\documentclass[fleqn,usenatbib,useAMS]{mnras}

\usepackage{graphicx}	
\usepackage{amsmath}	
\usepackage{amssymb}	
\usepackage{multicol}        
\usepackage{bm}		
\usepackage{pdflscape}	

\usepackage[T1]{fontenc}
\usepackage{ae,aecompl}

\usepackage{newtxtext,newtxmath}

\title[Galaxy populations from structured VAEs]{Structured Variational Inference for Simulating Populations of Radio Galaxies}

\author[D. J. Bastien et al.]{
David J. Bastien,$^{1,2}$\thanks{E-mail: david.bastien@manchester.ac.uk (DJB)}
Anna M. M. Scaife,$^{1,3}$
Hongming Tang,$^{1}$
Micah Bowles,$^{1}$
and 
Fiona Porter$^{1}$
\\
$^{1}$Jodrell Bank Centre for Astrophysics, Department of Physics \& Astronomy, University of Manchester, Oxford Road, Manchester M13 9PL, UK\\
$^{2}$Square Kilometre Array Organisation, Jodrell Bank Observatory, Macclesfield, SK11 9FT, UK\\
$^{3}$The Alan Turing Institute, Euston Road, London NW1 2DB, UK
}

\date{Accepted XXX. Received YYY; in original form ZZZ}

\pubyear{2021}

\begin{document}
\label{firstpage}
\pagerange{\pageref{firstpage}--\pageref{lastpage}}
\maketitle

\begin{abstract}
We present a model for generating postage stamp images of synthetic  Fanaroff-Riley Class~I and Class~II radio galaxies suitable for use in simulations of future radio surveys such as those being developed for the Square Kilometre Array. This model uses a fully-connected neural network to implement structured variational inference through a variational auto-encoder and decoder architecture. In order to optimise the dimensionality of the latent space for the auto-encoder we introduce the radio morphology inception score (RAMIS), a quantitative method for assessing the quality of generated images, and discuss in detail how data pre-processing choices can affect the value of this measure. We examine the 2-dimensional latent space of the VAEs and discuss how this can be used to control the generation of synthetic populations, whilst also cautioning how it may lead to biases when used for data augmentation.
\end{abstract}

\section{Introduction}

Since \cite{fanaroff1974} first introduced their classification scheme for radio galaxies, various works have focused on finding an explanation for such distinct morphological characteristics. \cite{ledlow1996} first showed that the FRI/II transition is both a function of the radio and optical luminosity. Other works focused on an explanation of the two groups through the underlying mechanism of the AGN and/or its environment. One of the most active debates considers the relationship between the accretion mechanism and the source morphology \citep{gendre2013}. Even if accretion correlation has not yet been established, \cite{croston2018} make a clear analysis of the two classes to conclude that the difference in external pressures of the lobes and jets, which is a distinctive intrinsic property to each classes, could be explained by the difference in the proton content causing a difference in energies in both classes. \cite{kaiser2007} worked on an evolutionary link between FRI and FRII, where they postulated that all radio galaxies must have at some point near the start of their life shown some FRII morphology. The lobes of sources with weak jets came into equilibrium with their surroundings causing the lobes to start close to the core. The jets are then disrupted by the ambient medium and the lobes eventually developed into FRIs.

However, the dividing line between FRI and FRII is blurry for low flux density samples. For different redshifts and environments, the morphology is sometimes uncertain and often leads to hybrid morphologies. \cite{best2009} identified sources with one lobe showing FRI-like morphology and the other showing FRII-like morphology. \cite{mingo2019} make use of a two-sided source classification method and found that $20 \%$ of their sources were of hybrid morphology.

It is evident that with new radio surveys this blurry line between FRI and FRII radio galaxies will be made clearer by (i) providing larger samples of sources at different redshifts and environment, and (ii) by providing larger samples of sources at low flux levels. Various research groups are thus working on surveys such as the Evolutionary Map of the Universe survey \citep[EMU;][]{norris2011} made using the Australian Square Kilometre Array Pathfinder (ASKAP), the MeerKAT International GHz Tiered Extra-galactic Exploration survey \citep[MIGHTEE;][]{jarvis2017} done using MeerKAT, the GaLactic and Extragalactic All-Sky MWA survey \citep[GLEAM;][]{hurley2016} done using the Murchison Widefield Array (MWA), the Very Large Array Sky Sky Survey \citep[VLASS;][]{villa2018} done using Very Large Array (VLA) and the latest LoTSS survey \citep{shimwell2019} done using LOFAR. 

With the advent of such huge surveys, new automated classification algorithms have been developed to replace the \emph{``by eye''} classification methods used in early classification. These new algorithms are trained using existing databases of pre-classified FRI and FRII sources. \cite{aniyan2017} made use of a deep convolutional neural network for the classification of FRI and FRII galaxies, while \cite{tang2019} made use of transfer learning to classify sources across different surveys introducing cross-survey identification abilities to existing neural networks. \cite{lukic2019} made use of capsule networks instead of CNNs for classifying images from the LOFAR Two-metre Sky Survey(LoTSS) survey. \cite{ma2019} trained an encoder-decoder architecture and used the encoder as a classifier, and finally \cite{wu2018} engineered the CLARAN network, which was able to locate and classify radio sources trained using the Radio Galaxy Zoo catalogue \citep{radiogalaxyzoo}.

In radio astronomy, deep learning has found its way through use in radio source classification in a similar way to classical ML. The ground work was done by \cite{aniyan2017} who made use of CNNs for the classification of radio galaxies. This was followed by other works involving the use of deep learning in source classification, \cite{lukic2018} made use of CNNs for the classification of compact and extended radio sources from the Radio Galaxy Zoo catalogue. CLARAN (Classifying Radio Sources Automatically with Neural Network) \cite{wu2018} made use of the Faster R-CNN \citep{2MASKRCNN} to identify and classify radio sources. \cite{alger2018} made use of an ensemble of classifiers including CNNs to perform host galaxy cross-identification. \cite{tang2019} made use of transfer learning with CNNs to do cross-survey classification. \cite{gheller2018} made use of deep learning for the detection of cosmological diffuse radio sources. \cite{lukic2019} performed morphological classification using a novel technique known as capsule networks. However, while most of these previous works focused on the use of classifiers, in this work we focus on the use of neural networks to generate realistic radio galaxies.

Source simulation plays an important role in the development of modern day telescopes. Simulated sources should reflect two very important characteristics: (i) They reflect the extrinsic properties of the astronomical entities being simulated. For example, if we are dealing with radio galaxies, the simulated sources should show the jets, lobes and core structure. (ii) They should simulate the properties of the telescopes, and as such simulate the bean width or artefacts that might result from any receiver characteristics. In optical astronomy, both aspects have been catered for by various authors and the importance of such simulations are well established.  For example, \cite{peterson2015} simulated astronomical objects for optical surveys and states the importance of such tools for the planning of future observations and for the development of sophisticated analysis software. 

In radio astronomy, equivalent simulators will be useful for the development of the SKA. Indeed, to test existing processing tools, in 2018 the SKA released the first SKA data challenge \citep{bonaldi2018}, which consisted of a series of generated images with specifications similar to those expected from the SKA. The community were invited to (i) undertake source finding on the generated images,  (ii) perform source property characterisation, and (iii) identify the different source populations. The final results from the challenge were bench-marked against the input catalogue used to generate the images. 

In addition to being used for bench-marking existing analysis tools, source simulators are also useful for improving existing classifiers. Most radio source classifiers make use of data augmentation to train their models, due to a lack of labelled training data (\cite{tang2019},\cite{wu2018},\cite{aniyan2017}). This augmentation involves rotating the images, flipping and inverting each image in the training set. However \cite{miko2018} concluded that even if they are fast and easy to implement, the data augmentation technique does not bring any new visual features to the data. In comparison, using simulated sources as training data might introduce new features and could improve the ability of a classifier. 

There are different methods that can be used to simulate radio sources. Radio Galaxies can be simulated through astrophysical fluid dynamics simulations. For example, \cite{rossi2017} made use of magneto-hydrodynamics (MHD) to simulate X-sources \citep{x_sources}, while \cite{tchekhovskoy2016} used MHD to simulate FRI and FRII jets. Although such simulations are highly realistic, they exhibit various practical issues when it comes to general use. Firstly, they do not cater for the telescope optics and are indeed designed to simulate perfect sources without those constraints. Secondly, when large populations of sources are needed this method turns out to be computationally inefficient and more complex models must be employed to cater for different environmental effects.
 
 \cite{bonaldi2018} made use of simple Gaussians to mimic radio sources in the SKA Challenge and \cite{makhathini2015} modeled radio galaxies by using triple Gaussians. However even if the generated images are not as realistic as real sources, they take into consideration the telescope characteristic. \cite{bonaldi2018} simulated images corresponding to the 3 frequencies of the SKA, namely 560\,MHz, 1.4\,GHz and 9.2\,GHz, also adding in noise and convolving the sources with the point spread function (PSF) of the telescope. However, even if such simulations respect telescope specifications, fine structures that are typically found in jets and lobes are absent. Hence an ideal simulator would be one which is able to generate images with the telescope properties \emph{and} generate structures similar to those made by MHD modelling. 
   
In this paper we use generative machine learning to simulate radio sources. Generative algorithms were first introduced in 2014 with two main methods: the Generative Adversarial Network \citep[GAN;][]{gan2014} and the Variational Autoencoder \citep[VAE;][]{kingma2014}. Generative algorithms are trained using real data to generate fake (or synthetic) data that look similar, without memorizing the training set. They have been employed across multiple different domains, including the generation of human faces \citep{karras2019},  the generation of artworks \citep{elgammal2017}, and to generate radio sources \citep{ma2018}.

The idea behind the use of neural networks for generating data is based on the ability of neural networks to reduce high dimensional data to lower dimensions or vice versa. These dimensional transforms can be performed using \emph{encoders} and \emph{decoders}. These two types of network are used in two basic applications of generative algorithms: generative adversarial networks (GANs) and variational auto-encoders (VAEs). VAEs makes use of both encoders and decoders. While encoders are not used in GANs, encoders do constitute an important building block for VAEs. Encoders work by reducing high dimensional data to lower dimensions. This is done by either using fully connected layers or CNNs. For VAEs, the encoder can be thought of as an embedding method similar to principal component analysis \citep[PCA; e.g.][]{vae_pca}. The decoder is the generative network common to all generative methods. Decoders use lower dimensional data to produce higher dimensional outputs. Similar to the encoder, this can done using a fully connected network or a de-convolution network.

These methods have been used used in both optical and radio regimes as a generative method, an unsupervised clustering tool and inference method. In the radio regime, \cite{ralph2019} made use of auto-encoders combined with self-organizing maps to separate outliers by making use of k-means clustering methods. \cite{regier} applied the VAE to astronomical optical images as an inference tool, while \cite{spindler2020} designed a VAE that can be used for unsupervised clustering and generation of synthetic optical images from the Sloan Digital Sky Survey (SDSS). \cite{ravan2016} make use of a conditional VAE for the generation of realistic galaxy images to be used as calibration data.

In this work, we use a variational auto-encoder (VAE) to implement a structured variational inference approach to generating simulated postage stamp images of radio galaxies in different target classes. In Section~\ref{sec:vae} we introduce variational auto-encoders and outline the statistical basis of their operation; in Section~\ref{sec:data} we introduce the data sets used for training the VAEs used in this work; in Section~\ref{sec:ramis_Score} we introduce the concept of the radio astronomy morphology inception score and define how it is calculated; in Section~\ref{sec:network} we describe the neural network implemented in this work and how the dimensionality of the latent space is optimised; in Section~\ref{sec:results} we give an analysis of the generated images from this network; in Section~\ref{sec:discussion} we discuss the implications of these results and how they differ from previous generative applications in the field; and in Section~\ref{sec:conclusion} we draw our conclusions.

\section{Variational Auto-Encoders}
\label{sec:vae}

The first proposed form for an auto-encoder was a neural network comprised of two fully-connected layers that transformed an input to an output without causing any distortion in the process \citep{rumelhart1985}. It did so by learning to generate a compact representation of the data, in a process similar to principal component analysis (PCA). This compact representation could then be used to re-generate the original input through the use of a decoder. Once trained, the decoder could be disconnected from the encoder to generate new data.  
In practice however, the original form of the encoder-decoder network was not able to create truly new data samples as the reduced space was not continuous. In order to address this problem of continuity, variational auto-encoders (VAEs) were proposed \citep{kingma2015}. The reduced space of the VAE is continuous by design, which implies that any point sampled from the reduced or \emph{latent} space can generate an output that will appear real. 

Since their first implementation by \cite{kingma2015}, VAEs have been used in many different fields, including video generation \citep{denton2018}, text generation 
\citep{semeniuta2017}, molecular design \citep{denovo2018} as well as in cosmology \citep{yi2020}. When it comes to improvements, various works have focused on the improvement of the original VAE. These include both domain specific VAEs and those that implement general improvements to the original architecture of \cite{kingma2015}. Among these are the infoVAE \citep{zhao2019}, the VAE-DGP \citep{dal2015}, the Hypersherical VAE \citep{davidson2018}, the WAE \citep{tolstikhin2017}, the control VAE \citep{shao2020}, the semi amortized VAE \citep{kim2018} and the $\delta$-VAE \citep{razavi2019}, which works to prevent posterior collapse in VAEs.

Arguably, the most major advances in the field of VAEs were made by \cite{kingma2014} and \cite{sohn2015}. \cite{kingma2014} improved their original VAE by engineering a semi-supervised VAE (sVAE) which incorporates a classifier and a VAE with conditioning. As such the networks could be trained using both unlabelled and labelled images. \cite{sohn2015} further modified the sVAE by removing the classifer and focused on the conditional properties of the training. The so-called Conditional VAE (CVAE) can be used to train a VAE to generate a specific target class, an approach that we follow in this work to generate FRIs and FRIIs, see Section~\ref{sec:cvae}. The CVAE has been used in various domains, such as dialog generation \citep{shen2017}, basketball player modelling \citep{acuna}, drug discovery \cite{polykovskiy2018}, sustainable building generation \citep{ge2019} and machine translation \citep{pagnoni2018}. Improvements were also made to the CVAE, such as the CDVAE \citep{lu2016}, the CF-VAE \citep{Bhat2018} and the W-VAE \citep{ryu2019}. 

Further improvements in VAEs were made by combining the VAEs with the discriminator of a Generative Adversarial Network (GAN). These VAEs are trained with a binary discriminator that classifies the generated images as fake or real. The VAEs are  trained so as to generate data that are classified as real by the discriminator. The VAE-GAN \citep{larsen2015} is a normal VAE fitted with a GAN discriminator. Similarly, the CVAE-GAN \citep{bao2017} and sVAE-GAN are CVAEs and sVAEs respectively, which are both fitted with a GAN discriminator.

To achieve a continuous latent space \cite{kingma2014} assumed that the latent space is represented by a continuous random variable with a parametric prior distribution. For neural networks that introduce non-linearity between their layers, calculation of the true posterior distribution for such a variable with respect to an input data set is generally intractable; however, the parameters of a variational approximation to the posterior can be learned instead.

\subsection{Variational Inference for Generative Models}
\label{sec:svi}

Consider that we have $N$ observable data points that are contained in a dataset $\mathbf{X}=\{ x^{(i)}\}_{i=1}^{N}$ where $\mathbf{x}^{(i)} \in \mathbb{R}^D$. We assume that these data points represent examples of a continuous random variable, $\mathbf{x}$, which can be encoded into a lower-dimensional latent space, $\mathbf{z}$, where $\mathbf{z} \in \mathbb{R}^d$ with $d<D$, and vice-versa that a sample drawn from \textbf{z} can be decoded to generate a new example, $\hat{\mathbf{x}}^{(i)} $. However, we do not \emph{a priori} know the distribution of $\mathbf{z}$ or the parameters of the function that maps one space to the other, $\mathbf{\theta}$. 

In cases where the encoder and decoder functions are complex and/or non-linear, both the true posterior, $p_{\mathbf{\theta}}(\mathbf{z}|\mathbf{x})$, and the marginal likelihood, $p_{\mathbf{\theta}}(\mathbf{x)}$, are typically intractable and therefore these relationships cannot be evaluated directly. In such cases a variational approximation to the true posterior is adopted and denoted $q_{\phi} (\textbf{z}|\textbf{x})$, where $\phi$ are the parameters of (e.g.) a neural network used to approximate the true transforms. Typically a Gaussian approximation to $p_{\mathbf{\theta}}(\textbf{z}|\textbf{x})$ is assumed and the Kullback-Leibler (KL) Divergence \citep{kullback1951} between $p_{\mathbf{\theta}}(\textbf{z}|\textbf{x})$ and $q_\phi (\textbf{z}|\textbf{x})$ is used as an optimization metric. 

Following \cite{kingma2014}, in this scenario the marginal log likelihood of an individual data point is given by,
\begin{equation}
\label{eu_eqn}
\log p_{\theta} (\mathbf{x}^{(i)}) = D_{KL}(q_{\phi} (\mathbf{z}|\mathbf{x}^{(i)}) || p_{\theta} (\mathbf{z}|\mathbf{x}^{(i)})) + \mathcal{L}(\mathbf{\theta}, \mathbf{\phi} ; \mathbf{x}^{(i)})
\end{equation}
where the first term is defined as,
\begin{equation}
\label{eq:kl}
D_{KL}(q_{\phi} (\textbf{z}|\textbf{x}^{(i)}) \Vert p_{\theta} (\textbf{z}|\textbf{x}^{(i)})) = - \sum  q_{\phi} (\textbf{z}|\textbf{x}^{(i)}) \log \frac {p_{\theta} (\textbf{z}|\textbf{x}^{(i)})}{q_{\phi} (\textbf{z}|\textbf{x}^{(i)})}
\end{equation}
and the second term is given by,
\begin{equation}
\label{eq:elbo}
    \mathcal{L}(\mathbf{\theta}, \mathbf{\phi} ; \mathbf{x}^{(i)}) = \mathbb{E}_{q_z(\mathbf{z} | \mathbf{x})} \left[ -\log q_z(\mathbf{z} | \mathbf{x}) + \log p_{\theta}(\mathbf{x}, \mathbf{z}) \right].
\end{equation}

Since Eq.~\ref{eq:kl} is always non-negative, this implies that $\log p_{\theta} (\mathbf{x}^{(i)}) \geq \mathcal{L}(\theta, \phi ; x^{(i)})$, which is referred to as the variational or evidence lower bound (ELBO) and is the loss term for training a VAE. 

In practice, the expression for the lower bound given in Equation~\ref{eq:elbo} can be re-written as:
\begin{eqnarray}
\label{eq:elbo2}
\nonumber \mathcal{L}(\mathbf{\theta}, \mathbf{\phi} ; \textbf{x}^{(i)}) = &-& D_{KL} (q_{\phi} (\textbf{z}|\textbf{x}^{(i)}) \Vert p_{\theta}(\textbf{z}))\\
&+& E_{q_{\phi} (\textbf{z}|\textbf{x}^{(i)})} [\log p_{\theta} (\textbf{x}^{(i)}|\textbf{z})]
\end{eqnarray}
and these two terms play a key role in the understanding of the general behaviour of the variational auto encoder. 

The first term in Equation~\ref{eq:elbo2} is the KL term and minimises the difference between the variational approximation and the true posterior. By rearranging Equation~\ref{eq:kl} and using the Gaussian re-parameterisation proposed by \cite{kingma2014}, which simplifies sampling a random deviate, $\tilde{z}$, from the Gaussian distribution, $\mathcal{N}(\mu, \sigma^2)$, by using the differentiable relationship
\begin{equation}
    \tilde{z} = \mu + \sigma \cdot \epsilon~~{\rm where}~~\epsilon \sim \mathcal{N}(0,1),
\end{equation}
and we can therefore rewrite the KL term as:
 \begin{equation}
 \label{eq:term1}
     D_{KL} (q_{\phi} (\textbf{z}|\textbf{x}^{(i)}) \Vert p_{\theta}(\textbf{z})) = \frac{1}{2} \sum_{j=1}^d (1 + \log \sigma_j^2 -\mu_j^2-\sigma_j^2),
 \end{equation}
where $p_{\theta}(\textbf{z}) = N(0,1)$ and $q_{\phi}(\textbf{z}|\textbf{x}^{(i)})= N(\mu,\sigma^2)$.

The second term in Equation~\ref{eq:elbo2} can be considered as the reconstruction error and ensures that the network correctly reconstructs the input image. Since $p_{\theta}(\textbf{x}|\textbf{z})$ is a function that maps $\textbf{z}$ into $\hat{\textbf{x}}$, where $\hat{\textbf{x}}$ represents the reconstructed image, we can assume that the term will take a form $p_{\theta}({\textbf{x}}|\hat{\textbf{x}})$ and in the Gaussian case,
 \begin{equation}
 \label{eq:term2}
     E_{q_{\phi} (\textbf{z}|\textbf{x}^{(i)})} [\log p_{\theta} (\textbf{x}^{(i)}|\textbf{z})] = \frac{1}{N} \sum_{l=1}^{N} ({\textbf{x}^{(l)}}-\hat{\textbf{x}}^{(l)})^2.
 \end{equation}

Using Equations~\ref{eq:term1}~\&~\ref{eq:term2}, we can therefore express the full loss term as:
\begin{equation}
\mathcal{L}(\theta, \phi ; \textbf{x}^{(i)}) = - \frac{1}{2} \sum_{j=1}^d (1 + \log \sigma_j^2 -\mu_j^2-\sigma_j^2) + \frac{1}{N} \sum_{l=1}^{N} ({\textbf{x}^{(l)}}-\hat{\textbf{x}}^{(l)})^2.
\label{eqn:loss_function}
\end{equation}

Since the marginal log likelihood of the full training data set is given by the sum over the marginal log likelihoods of each individual data point, the total loss for the full data set is therefore given by
\begin{equation}
\mathcal{L}(\theta, \phi ; \textbf{x}) = \sum_{i=1}^{N} \mathcal{L}(\theta, \phi ; \textbf{x}^{(i)}).
\end{equation}

\subsection{Conditional VAEs}
\label{sec:svi_cvae}

In the conditional case we again have $N$ observable datapoints that are represented by the dataset $\hat{\textbf{X}}=\{ \textbf{x}^{(i)}\}$, where $\textbf{x}^{(i)} \in R^{D}$. But it is now accompanied by the associated labels or class condition,  given as $\hat{\textbf{Y}}=\{ \textbf{y}^{(i)}\}$. Similarly the data will be generated by the latent space $z$ where $\textbf{z}^{(i)} \in R^{d}$ where $d<D$. The encoder will take as input $\textbf{X}$ and $\textbf{Y}$ to encode $z$ and the decoder will take $z$ and $\textbf{Y}$ to generate the data point $\hat{\textbf{x}}^{(i)}$. We here make use of Bayes Theorem with conditioning that takes into consideration the class condition, $y$,
\begin{equation}
p_{\theta}(\mathbf{z}|\mathbf{x},\textbf{y})=\frac{p_{\theta}(\mathbf{x}|\mathbf{z},\textbf{y})p_{\theta}(\mathbf{z}|\textbf{y})}{p_{\theta}(\mathbf{x}|\textbf{y})}.
\end{equation}

In this case the marginal likelihood of an individual data sample can be written as,
\begin{equation}
\log p_{\theta} (\mathbf{x}^{(i)},\textbf{y}) = D_{KL}(q_{\phi} (\mathbf{z}|\mathbf{x}^{(i)},\textbf{y}) || p_{\theta} (\mathbf{z}|\mathbf{x}^{(i)}, \textbf{y})) + \mathcal{L}(\theta, \phi ; x^{(i)},\textbf{y})
\end{equation}
and consequently the ELBO is defined as,
\begin{eqnarray}
\label{eq:elbocvae}
\nonumber \mathcal{L}(\mathbf{\theta}, \mathbf{\phi} ; \mathbf{x}^{(i)},\textbf{y}) = &-&  D_{KL}(q_{\phi} (\textbf{z}|\textbf{x}^{(i)},\textbf{y}) \Vert  p_{\theta} (\textbf{z}|\textbf{y}))\\
&+&  \mathbb{E}_{q_{\phi} (\textbf{z}|\textbf{x}^{(i)},\textbf{y})} [\log p_{\theta} (\textbf{x}^{(i)}|\textbf{z},\textbf{y})].
\end{eqnarray}




\section{Data Set Definitions}
\label{sec:data}

In this work, we make use of FRDEEP, first introduced in \cite{tang2019}. The FRDEEP data set consists of an imbalanced sample of FRI and FRII radio galaxy images, extracted from the FIRST radio survey \citep{becker} and subjected to image pre-processing using the method described in \cite{aniyan2017}. It makes use of two catalogues (i) the combined NVSS and FIRST Galaxies Catalog  \citep[CoNFIG;][]{gendre2013,gendre2010} and (ii) the FRICAT catalogue \citep{FRICAT} to label the images. Here we describe the data set in more detail.

\subsection{The combined NVSS and FIRST Galaxies Catalog (CoNFIG)}
The CoNFIG catalogue of FRI and FRII galaxies is constructed from 4 sub-samples referred to as ConFIG-1, 2, 3 and 4. These subsamples contain sources that were originally selected from the NVSS survey \citep{condon1998} with flux density cuts of $S_{\rm 1.4\,GHz} \geq 1.3, 0.8, 0.2$ and $0.005$\,Jy. These sources were first identified using the NVSS catalogues using the process brought forward by \cite{gendre2008} where they make use of component association to identify the radio galaxies. If two or more components were found with $S_{\rm 1.4\,GHz} > 1.3$\,Jy, the components were considered as being part of a larger resolved source. For multiple component sources with $0.1 \geq S_{\rm 1.4\,GHz} \geq 1.3$\,Jy located within 4\,arcmin of the listed sources were set aside for visual inspection. The identified sources were classified as FRI and FRII either by referring to existing literature or by visually classifying them. Once identified, the sources were visually classified. The sources were classified as FRII if distinct hotspots were displayed at the edge of the lobes and that the lobes we aligned. If collimated jets were observed and hot-spots were observed close to the core and jets, these sources were further classified as FRI.

If the sources were smaller than 1 arcsec, they were classified as compact. The sources which had an extended morphology were classified as FRI and FRII. Using the above criteria 849 sources were identified. These sources were classified as FRI and FRII, compact (with no classification) or uncertain. The sources were further tagged as 'confirmed' or 'possible'. This resulted to a final catalogue that contains 50 confirmed FRI and 390 confirmed FRII.

\subsection{FRICAT}
To balance the ConFIG dataset that consists mostly of FRII sources, we make use of the FRICAT. The catalogue consists of 219 sources that were first identified using the database of  \cite{best2012} identified using the method brought forward by \cite{best2009}. The initial sample consists of 18,286 radio galaxies. These sources were cross-matched with the optical spectroscopic data from the data release 7  of the Sloan Digital Sky Survey (SDSS). They perform a redshift selection within $z < 0.15$ where 3357 sources were selected. The FIRST sources were then visually inspected and the sources within an extent of 30 kpc from the optical host were chosen, this corresponds to 11.5\,arcsec and is ideal for morphological classification with a 5$''$ resolution.
 
 The sources were then visually classified, they limited their selection to one-side and two-sided jets and focused on those sources whose brightness decreased along the jets with no enhanced brightness at the jet end. The classification was performed independently by the three authors and accepted if two authors agreed on a classification. The final catalogue consists of 219 FRI sources.

\subsection{Combined Data}

Out of the combined 659 sources from CoNFIG and FRICAT, 600 sources were randomly selected to reduce class imbalance, of which 264 sources were FRIs and 336 were FRIIs. The training set consists of 550 sources and 50 sources were assigned to the test set. Once augmented, the training set consists of 198,000 sources, of which 87,120 are FRIs and 198,000 are FRIIs. Our source selection is shown in Table \ref{tab:source_selection} and the data pre-processing and augmentation process is explained in the following section.
\begin{table}
\centering
\begin{tabular}{lccc}
\hline
                                      & \textbf{FRI} & \textbf{FRII} & \textbf{Total} \\ \hline
\textbf{FRICAT}                       & 219          & -             & 219            \\
\textbf{ConFIG}                       & 50           & 390           & 440            \\
\textbf{Total \# of Sources}          & 269          & 390           & 659            \\ \hline
\textbf{Total \# of Selected Sources} & 264          & 336           & 600            \\ \hline
{Training}          & 242          & 308           & 550            \\
{Test}           & 22           & 28            & 50             \\ \hline
\textbf{Train Augmented}              & 87,120       & 110,880       & 198,000         \\ \hline
\end{tabular}
\caption[Number of selected FRI and FRII sources selected from the FRICAT and ConFIG] { Number of selected FRI and FRII sources selected from the FRICAT and ConFIG. 600 sources were selected from both datasets with 550 Train sources and 50 test sources. }
\label{tab:source_selection}
\end{table}

\subsection{Image Pre-processing and Data Augmentation}
\label{sec:image_aug}

An important procedure in machine learning is data pre-processing, which helps to maintain a homogeneous sample space \citep{aniyan2017}. While human classifiers can easily deal with the background noise in images and classify objects oriented differently, ML classifiers perform badly  if an attempt is made to classify radio sources without a proper pre-processing procedure. For CNNs, \cite{aniyan2017} and \cite{tang2019} have shown that the noise should first be clipped at 3-sigma level, the pixels values rescaled between 0 and 1 and finally the images augmented through flipping and rotation for the classifier to attain a good accuracy. \cite{wu2018} applied a different pre-processing procedure, they performed a zero-centering of the pixel values followed by a rescaling of the source followed by a horizontal flip to attain good performance. Our image pre-processing makes use of the method used by \cite{aniyan2017} and \cite{tang2019}. This was done in two phases. The first phase involves the processes covered by \cite{tang2019} where the pre-processed data was made available through their git repository\footnote{\url{https://github.com/HongmingTang060313/FR-DEEP}}. This data was reprocessed to adapt to our needs in the VAE. To augment the training dataset the sources were rotated from 0 to 360 degrees in intervals of 1 degree. This resulted in a total of 198,000 sources in the training set with 87,120 FRI sources and 110,880 FRIIs. Table~\ref{tab:source_selection} summarises this source distribution.

\section{Radio Astronomy Morphology Inception score}
\label{sec:ramis_Score}

One of the main challenges of developing and implementing generative algorithms is that they are difficult to evaluate and compare. Compared to discriminative methods, where we can make use of metrics like accuracy, the F1-Score, the precision or the recall, generative algorithms do not have such direct measures for comparing different outcomes. While the loss can be used as an indication of the model performance, it cannot be used comparatively across different architectures or algorithms.

To deal with this issue, we make use of an adapted form of the \emph{inception score} \citep{inception2014}, a measure that can be used for comparison across different generative algorithms. This standardized method makes use of classifiers to quantify the generated image quality by evaluating the degree of uncertainty in its classification. The inception score was first introduced using the inception v1 model, trained on 1000 target classes of the Imagenet Large Scale Visual Recognition (ILSVRC) 2014 classification challenge \citep{ILSVRC15}. The inception classifier was used to evaluate the uncertainty in image classification: images that were correctly generated were assigned to one of the 1000 classes and assigned a high probability, while those that were incorrectly generated were assigned to multiple classes with low probability. The inception score was then used as a measure of this difference.

\begin{figure*}
\centering
\includegraphics[width=1.0\textwidth]{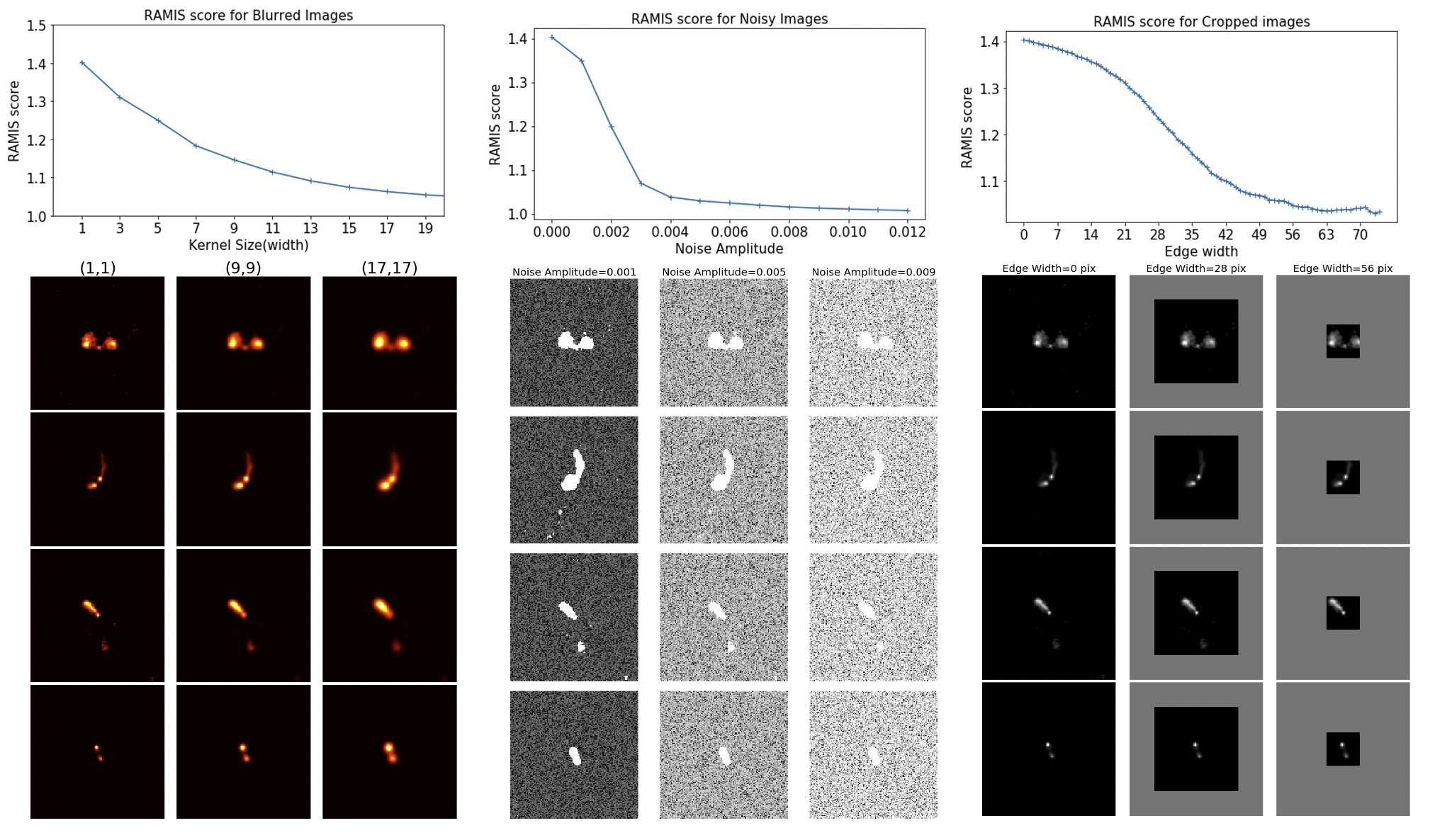}
\caption[Experiment 1: Image blurring]{Experiment 1,2 and 3:Experiment 1 involves the image blurring with kernel size (1,1) (No blurring) to (19,19). Experiment 2 involves the cropping of the images while experiment 3 involves the addition of random noise.}
\label{fig:experiment_blurry}
\end{figure*}

To calculate the score, the inception model was applied to each generated image and the conditional probability, i.e. the probability that the image belongs to one class, $p(y|x)$, was obtained. Using those probabilities, the entropy was calculated in order to evaluate the inception score, which is given by the KL divergence between $p(y|x)$ and $p(y)$:
\begin{equation}
              I_{\rm score} = \exp \left[D_{\rm KL}(p(y|x)||p(y) \right].
              \label{eqn:inception_Score}
    \end{equation}
Both $p(y|x)$ and $p(y)$ are evaluated from the generated dataset: $p(y|x)$ is the output of the classifier, i.e. the label distribution of an image where $y$ is the set of labels and $x$ is the image; $p(y)$ is the marginal distribution of the labels, $y$, for the generated dataset.  

A high inception score implies a well performing generative algorithm and the  highest inception score to date for the ILSVRC~2014 classification data set is 9.46, based on the de-noising diffusion probabilistic mode \citep{ho2020}. 

Although we wish to use the concept of the inception score to evaluate radio source generation, the inception model itself was not trained using radio sources. To address this we make use of the existing radio source classifier architecture introduced in \cite{tang2019}, which uses an architecture similar to \cite{aniyan2017}. This classifier consists of 5 convolutional layers each with batch normalisation and max pooling. The output of the final convolution layer is flattened and input into a fully connected network consisting of 3 layers. The output from the fully connected layers is then passed through a softmax function to obtain the predictions $p(y|x)$.

This CNN was trained using the FRDEEP dataset described in Section~\ref{sec:data} using 390 sources, validated using 110 sources, and tested using 55 sources. The data was augmented and pre-processed using the method described in Section~\ref{sec:image_aug}. We make use of the Adagrad optimizer with an initial learning rate of 0.001. The network was trained for 30 epochs and the training was stopped when the validation and training loss stabilized. Using this network, we are able to evaluate $p(y|x)$ and calculate the RAdio Morphology Inception Score (RAMIS), our equivalent of the original inception score for radio astronomy. 

\subsection{Metric Performance Evaluation}
\label{sec:ramis_performance}

To test the ability of the RAMIS to quantify the quality of generated images, we performed a number of tests, transforming the training images from FRDEEP in different ways and evaluating the RAMIS score for each transformation. To perform these tests, we make use of the non-augmented training data set from FRDEEP. These images once transformed were passed through the classifier model to obtain $p(y|x)$ and evaluate the RAMIS score using Equation~\ref{eqn:inception_Score}. 

The original images with no transformation applied had a RAMIS score of 1.4 when evaluated. We then performed the following transformations on those images, see Figure \ref{fig:experiment_blurry} : (i) blurring of the images using kernels of different sizes; (ii) addition of noise at varying levels; and (iii) cropping of the images.

In the first  experiment, we performed a Gaussian blurring and vary the size of the smoothing kernel from $(1,1)$ to $(19,19)$ pixels in steps of 2. This results in a decrease in the resolution of the images. The RAMIS score was evaluated for each kernel size. 

An exponential drop in the RAMIS was observed from 1.4 to 1.05 where the non-transformed images with a kernel size of (1,1) resulted in a RAMIS of 1.4. The classifier confusion is similar to that obtained when observing at different resolving power, for example in Figure \ref{fig:experiment_blurry} row 3 the core and the jets are clearly unresolved, which causes class confusion within the model and reduces the RAMIS.

The second experiment involved the addition of random noise to the image, see Figure~\ref{fig:experiment_blurry}. We added random noise with varying amplitude between 0 and 0.011 in steps of 0.01 and evaluated the RAMIS at each stage. With no added noise the RAMIS score was 1.4, and a drastic drop in the score is observed from noise level 0.001 to 0.003. Above that noise threshold the classifier was strongly influenced by the noise and eventually randomly classifies the images.

The final experiment involved cropping the radio images to different sizes. In Section~\ref{sec:image_aug}, where we cover the pre-processing process, we cropped the input images to 100 x 100 pixels, i.e blanked an edge strip with a width of 25 pixels. This was done to prevent corner differences in the images when rotating. This process may remove information from some images, for example those that include low level features that might result from lobes or jets. To quantify this effect, the training images were cropped by blanking the edge with strips of width 0 to 75 pixels, at which point the image is completely blank. In a similar manner to the previous experiments, the RAMIS dropped exponentially as a function of strip width, converging towards 1.0 (worst score). At a strip width of 25 pixels, which corresponds to the pre-processing procedure used in this work, a RAMIS of 1.25 was measured.

This analysis is relevant when quantitatively evaluating the performance of VAEs or other generative algorithms. For example \cite{ma2019} cropped their input images to 40 x 40, corresponding to an edge blanking of 55 pixels. At that level the RAMIS is less than 1.05. This may be explained by the fact that many FRII lobes, which range in radial distance up to 75 pixels from the image centre, have been cut out of the images, causing only the core of the radio galaxy to be seen by the VAE. This will result in poorer performance when generating FRIIs. 

\section{Network Architecture \& Implementation}
\label{sec:network}

The unsupervised VAE used in this work consists of two fully connected neural networks: (i) the encoder and (ii) the decoder.  While many VAE implementations make use of convolutional layers \citep[e.g.][]{ma2018,ralph2019,spindler2020}, in this work we choose to use a fully connected network. There are a number of known issues involved in the use of convolutional VAEs and these are described in more detail in Section~\ref{sec:discussion}. Consequently, although the use of convolutional layers may be addressed in future work, here we retain a fully-connected architecture for simplicity.

\begin{figure*}
\includegraphics[width=0.92\textwidth]{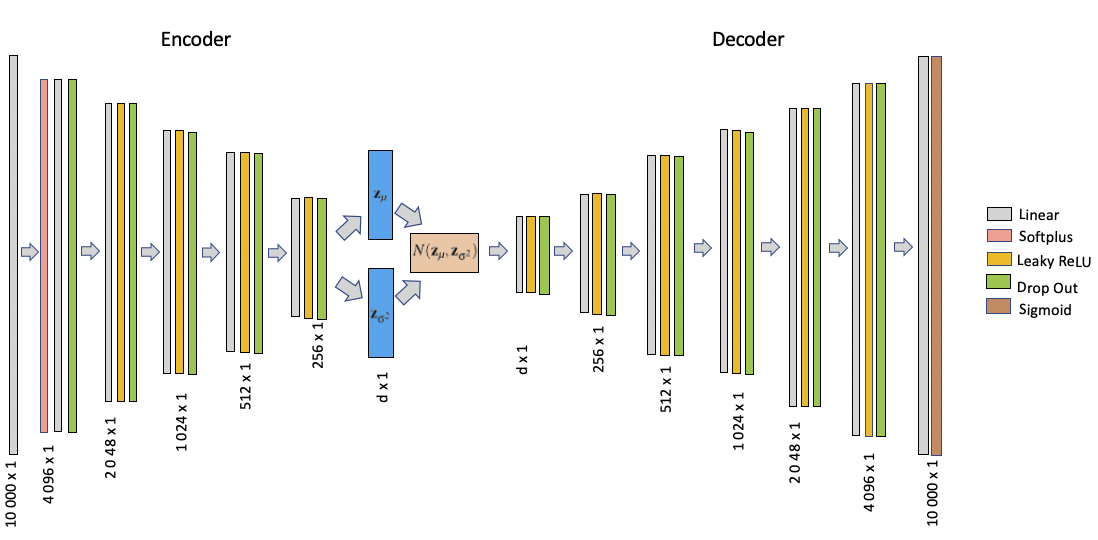}
\caption[Encoder-Decoder Fully Connected Architecture]{The VAE neural network architecture: The encoder reduces the dimensions of the input images through the use of a funnel architecture that halves at every layer in the encoder and doubles for the decoder.\label{fig:funnel}}
\end{figure*}

The encoder takes in the $100 \times 100$ pixel input images that have been reshaped into a $10000 \times 1$ vector using an input layer of the same dimension. We then make use of a funnel architecture that reduces the high dimensional inputs, $x$, to a lower latent dimension, $z$, where $z \in R^d$ and $x \in R^D$ with $D>d$, see Figure~\ref{fig:funnel}.
\begin{figure*}
\centering
\includegraphics[width=0.9\textwidth]{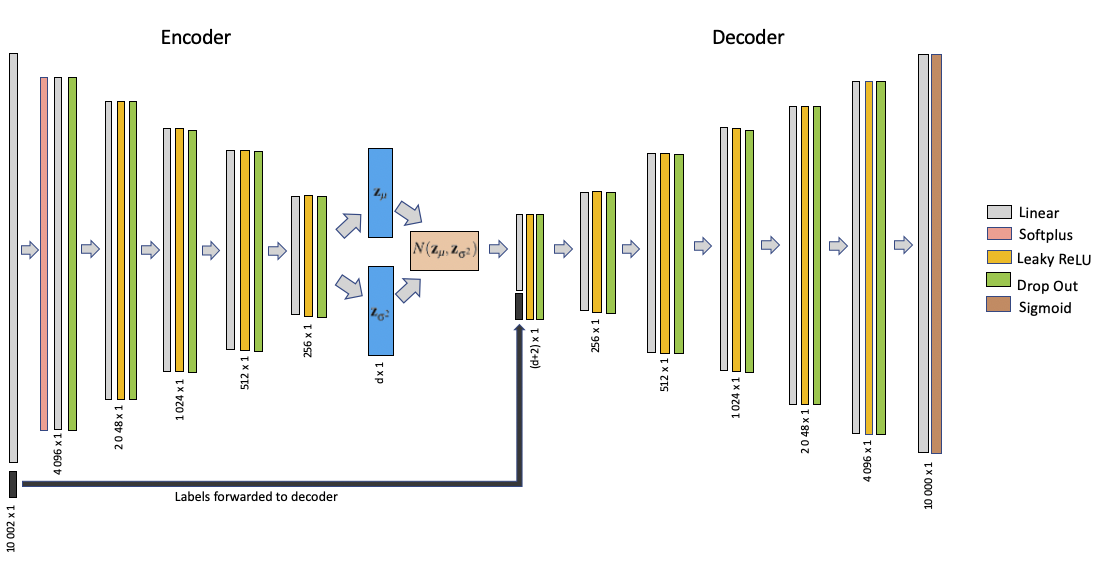}
\caption[Encoder-Decoder Fully Connected Architecture]{Modifications to the VAE for the CVAE model with two additional neurons at the input layer of the encoder and decoder. Labels are forwarded at both the encoder and decoder. \label{fig:funnel2}}
\end{figure*}
The first hidden layer has a soft-plus activation function and is fitted with a drop-out. The number of neurons are halved for each subsequent layer so as to create the tunnel architecture with the 2nd layer consisting of 2048 neurons, 3rd layer - 1024 neurons, 4th layer - 512 neurons and 6th layer - 256 neurons. Each layer was fitted with a leaky ReLU. As compared with the widely used Rectified Linear Unit (ReLU) which is a non-linear activation function that allows positive outputs from the neuron to pass while zeroing any negative values given by the function:
          \begin{equation}
            \begin{aligned}
            f(x)=\begin{cases} x & \text{if $x>0$} \\
            0 & \text{if $x \leq 0$}. \\
            \end{cases}
            \end{aligned}
            \end{equation}
The leaky ReLU resolves the inability of the ReLU to map the negative values by introducing a small negative slope to any negative input by following the function:
          \begin{equation}
            \begin{aligned}
            f(x)=\begin{cases} x & \text{if $x>0$} \\
            \alpha x & \text{if $x \leq 0$} \\
            \end{cases}
            \end{aligned}
            \end{equation}
Where $\alpha = 0.001$ in our work. All these hidden layers are fitted with a Leaky ReLU activation function and followed by a drop-out with $p=0.2$. The final output layer of the encoder consists of two $d$ dimensional layers where one layer outputs the mean parameters and the other layer outputs the variance parameter. 

The decoder on the other hand converts the sampled $z$ values into outputs with the same dimension as the encoder input. The decoder input takes the latent variable, $z$, that is sampled from the two Gaussians with parameters $z_{\mu}$ and $z_{\sigma^2}$. This sampling  The first layer is fitted with a Leaky ReLU and a drop-out. The network is then a mirrored version of the encoder where the 2nd layer has 256 neurons, 3rd layer - 2048 neurons, 4th - 1024 neurons, 5th - 2048 neurons and finally 6th - 4056 neurons. These layers are each fitted with a leaky ReLU and drop-out. The final layer is followed by a sigmoid activation that bounds the output between 0 and 1.

\begin{table*}
\centering

\begin{tabular}{cclcclcclcc}
\cline{1-5} \cline{7-11}
\multicolumn{5}{c}{\textbf{VAE Architecture}}                                                                                                                                                                                                         &  & \multicolumn{5}{c}{\textbf{CVAE Architecture}}                                                                                                                                                                         \\ \cline{1-5} \cline{7-11} 
\multicolumn{2}{c}{\textbf{Encoder}}                                                                                              &  & \multicolumn{2}{c}{\textbf{Decoder}}                                                                           &  & \multicolumn{2}{c}{\textbf{Encoder}}                                                               & \multicolumn{1}{c}{} & \multicolumn{2}{c}{\textbf{Decoder}}                                                       \\ \cline{1-2} \cline{4-5} \cline{7-8} \cline{10-11} 
\textbf{Layer}                                        & \textbf{\begin{tabular}[c]{@{}c@{}}Dimensions/\\ parameters\end{tabular}} &  & \textbf{Layer} & \multicolumn{1}{l}{\textbf{\begin{tabular}[c]{@{}l@{}}Dimensions/\\ Parameters\end{tabular}}} &  & \textbf{Layer}         & \textbf{\begin{tabular}[c]{@{}c@{}}Dimensions/\\ parameters\end{tabular}} & \multicolumn{1}{c}{} & \textbf{Layer} & \textbf{\begin{tabular}[c]{@{}c@{}}Dimensions/\\ Parameters\end{tabular}} \\ \cline{1-2} \cline{4-5} \cline{7-8} \cline{10-11} 
Input Layer                                           & $10,000 \times 1$                                                         &  & Linear FC      & $d \times 1$                                                                                  &  & Input Layer            & \begin{tabular}[c]{@{}c@{}}$(10,000+2) \times$ \\ $1$\end{tabular}        &                      & Linear FC      & $(d+2) \times 1$                                                          \\ \cline{1-2} \cline{7-8}
Softplus                                              & -                                                                         &  & Leaky ReLU     & -                                                                                             &  & Softplus               & -                                                                         &                      & Leaky ReLU     & -                                                                         \\
Linear FC                                             & $4,096 \times 1$                                                          &  & Dropout        & $p=0.2$                                                                                       &  & Linear FC              & $4,096 \times 1$                                                          &                      & Dropout        & $p=0.2$                                                                   \\ \cline{4-5} \cline{10-11} 
Dropout                                               & $p = 0.2$                                                                 &  & Linear FC      & $256 \times 1$                                                                                &  & Dropout                & $p=0.2$                                                                   &                      & Linear FC      & $256 \times 1$                                                            \\ \cline{1-2} \cline{7-8}
Linear FC                                             & $2,048 \times 1$                                                          &  & Leaky ReLU     & -                                                                                             &  & Linear FC              & $2,048 \times 1$                                                          &                      & Leaky ReLU     & -                                                                         \\
Leaky ReLU                                            & -                                                                         &  & Dropout        & $p=0.2$                                                                                       &  & Leaky ReLU             & -                                                                         &                      & Dropout        & $p=0.2$                                                                   \\ \cline{4-5} \cline{10-11} 
Dropout                                               & $p = 0.2$                                                                 &  & Linear FC      & $512 \times 1$                                                                                &  & Dropout                & $p=0.2$                                                                   &                      & Linear FC      & $512 \times 1$                                                            \\ \cline{1-2} \cline{7-8}
Linear FC                                             & $1,024 \times 1$                                                          &  & Leaky ReLU     & -                                                                                             &  & Linear FC              & $1,024 \times 1$                                                          &                      & Leaky ReLU     & -                                                                         \\
Leaky ReLU                                            & -                                                                         &  & Dropout        & $p=0.2$                                                                                       &  & Leaky ReLU             & -                                                                         &                      & Dropout        & $p=0.2$                                                                   \\ \cline{4-5} \cline{10-11} 
Dropout                                               & $p=0.2$                                                                   &  & Linear FC      & $1,024 \times 1$                                                                              &  & Dropout                & $p=0.2$                                                                   &                      & Linear FC      & $1,024 \times 1$                                                          \\ \cline{1-2} \cline{7-8}
Linear FC                                             & $512 \times 1$                                                            &  & Leaky ReLU     & -                                                                                             &  & Linear FC              & $512 \times 1$                                                            &                      & Leaky ReLU     & -                                                                         \\
Leaky ReLU                                            & -                                                                         &  & Dropout        & $p=0.2$                                                                                       &  & Leaky ReLU             & -                                                                         &                      & Dropout        & $p=0.2$                                                                   \\ \cline{4-5} \cline{10-11} 
Dropout                                               & $p=0.2$                                                                   &  & Linear FC      & $2,048 \times 1$                                                                              &  & Dropout                & $p=0.2$                                                                   &                      & Linear FC      & $2,048 \times 1$                                                          \\ \cline{1-2} \cline{7-8}
Linear FC                                             & $256 \times 1$                                                            &  & Leaky ReLU     & -                                                                                             &  & Linear FC              & $256 \times 1$                                                            &                      & Leaky ReLU     & -                                                                         \\
\begin{tabular}[c]{@{}c@{}}Leaky\\  ReLU\end{tabular} & -                                                                         &  & Dropout        & $p=0.2$                                                                                       &  & Leaky ReLU             & -                                                                         &                      & Dropout        & $p=0.2$                                                                   \\ \cline{4-5} \cline{10-11} 
Dropout                                               & $p=0.2$                                                                   &  & Linear FC      & $4,096 \times 1$                                                                              &  & Dropout                & $p=0.2$                                                                   &                      & Linear FC      & $4,096 \times 1$                                                          \\ \cline{1-2} \cline{7-8}
$z_{\mu},z_{\sigma^2}$                                & $d \times 1$                                                              &  & Leaky ReLU     & -                                                                                             &  & $z_{\mu},z_{\sigma^2}$ & $d \times 1$                                                              &                      & Leaky ReLU     & -                                                                         \\ \cline{1-2} \cline{7-8}
                                                      &                                                                           &  & Dropout        & $p=0.2$                                                                                       &  & \multicolumn{1}{l}{}   & \multicolumn{1}{l}{}                                                      &                      & Dropout        & $p=0.2$                                                                   \\ \cline{4-5} \cline{10-11} 
                                                      &                                                                           &  & Linear FC      & $10,000 \times 1$                                                                             &  & \multicolumn{1}{l}{}   & \multicolumn{1}{l}{}                                                      &                      & Linear FC      & $10,000 \times 1$                                                         \\
                                                      &                                                                           &  & Sigmoid        & -                                                                                             &  & \multicolumn{1}{l}{}   & \multicolumn{1}{l}{}                                                      &                      & Sigmoid        & -                                                                         \\ \cline{1-2} \cline{4-5} \cline{7-8} \cline{10-11} 
\end{tabular}

\caption{VAE and CVAE architectures.}
\label{tab:vae_cav_architecture}
\end{table*}

The architecture of the encoder-decoder network is shown in Figure~\ref{fig:funnel} and detailed in Table~\ref{tab:vae_cav_architecture}.

For the declarative definition. our encoder and decoder were defined within the model and guide as:
\begin{itemize}
    \item $p_{\theta}(z) = N(0,I)$
    \item $p_{\theta}(x|z) = H_{\theta} (z)$ where $H_{\theta}(z)$ represents the encoder
\end{itemize}
The guide which was introduced in section \ref{sec:svi} is given as:
\begin{itemize}
    \item $q_{\phi}(z|x) = N(z_{\mu},z_{\sigma^2})$ where $z_{\mu}=F_{\phi}(x)$ and $z_{\sigma^2}=G_{\phi}(x)$ where $F_{\phi}$ and $G_{\phi}$ are the same neural network with the final output layer outputs $z_{\mu}$, the mean parameters and $z_{\sigma^2}$, the variance parameters.
\end{itemize}
The VAE is trained by optimizing the guide to match the model so as to minimize the loss function derived in equation \ref{eqn:loss_function}. 

\subsection{Conditional VAE}
\label{sec:cvae}

The conditional VAE is a variation on the unsupervised VAE that also takes in the labels on the data. The labels are the conditions that associate particular data samples and can be used to generate images based on specific conditions. We input these labels as one hot-vectors at two instances in the network: firstly at the input to the encoder along with the reshaped images, and secondly at input to the decoder along with the latent $z$ for image reconstruction. The alterations to the unsupervised VAE are shown in figure \ref{fig:funnel2} and table \ref{tab:vae_cav_architecture}.\\
The model was also modified following the method outlined in Section ~\ref{sec:svi_cvae}, to accommodate this additional information. We alter the definition of the model to include a prior on the class, $p(y)=cat(y|\pi)$, and we alter the likelihood, $p_{\theta}(x|z)$, to be $p_{\theta}(x|z,y)={\rm Bernouilli}(x|H_{\theta}(z,y))$, where $H_{\theta}$ represents the encoder. The variational approximation, or guide, remains unchanged with $q_{\phi}(z|x) = N(z_{\mu},z_{\sigma^2})$, where $z_{\mu}=F_{\phi}(x,y)$ and $z_{\sigma^2}=G_{\phi}(x,y)$.

All networks used in this work were implemented using the Python probabilistic programming library {\tt Pyro} \citep{pyro1,pyro2}\footnote{Code for this work is available on the git repository: \url{https://github.com/joshen1307/RAGA}}.

\subsection{Learning Rate Search and Training}
The unsupervised VAE was first trained by performing a learning rate search. The learning rate search was performed for the different latent dimensions $d=2,4,8,16,32$ and $64$. This was done to identify the ideal learning rate that leads to the lowest train loss after 10 epochs. The VAE was trained using initial learning rates between 0.0005 to 0.0015 in steps of 0.0001. Such a procedure is crucial as the initial learning rate is often considered to be the most important hyper-parameter \citep{smith2017}. Table \ref{tab:vae_latent_dimension} shows the ideal initial learning rate for each latent dimension $d$.

The identified initial learning rates were used to train the network for 8000 epochs. For all the latent dimensions, the loss converged towards a minimum showing an adequate convergence towards a minimized loss. However the minimum differs for the different latent dimensions. None of the networks over-fitted the training data and the test loss remained stable after 6000 epochs. At $d=8$, the test loss stabilizes at $~130$. For the other latent dimensions i.e $d=2,4,16,32$ and $64$, the test loss stabilizes at $~120$. This difference in test loss had an impact on the RAMIS score across the epochs. Table \ref{tab:vae_latent_dimension} and figure \ref{tab:vae_latent_dimension} shows the train loss, test loss and RAMIS for the different latent dimensions.

A similar two phase procedure was adopted for the training of the Conditional VAE. We again made use of a learning rate search for the different latent dimensions $d=2,4,8,16,32$ and $64$. Using the identified learning rate, the CVAE was trained for each selected latent dimensions ( as shown in table \ref{tab:CVAE_latent_dimension}). In parallel to the VAE, the initial learning rate search was done between 0.0005 to 0.0015 in steps of 0.0001. Once identified, the CVAE was trained for 5000 epochs, where the test loss, train loss and RAMIS score was calculated every 10 epochs. In addition to those three metrics, we also generated 100 FRIs and 100 FRIIs which were classified using the classifier described in section \ref{sec:ramis_Score} . This was done to find the class generation efficiency and to identify any class bias in our CVAE. Those results are shown in table \ref{tab:CVAE_latent_dimension} and figure \ref{fig:CVAE_latent_dimension}, the additional metrics were denoted as $FRI_{count}$ and $FRII_{count}$ are the respective fraction of generated FRI and FRII that were correctly classified.

\begin{figure}
\centering
\includegraphics[width=0.5\textwidth]{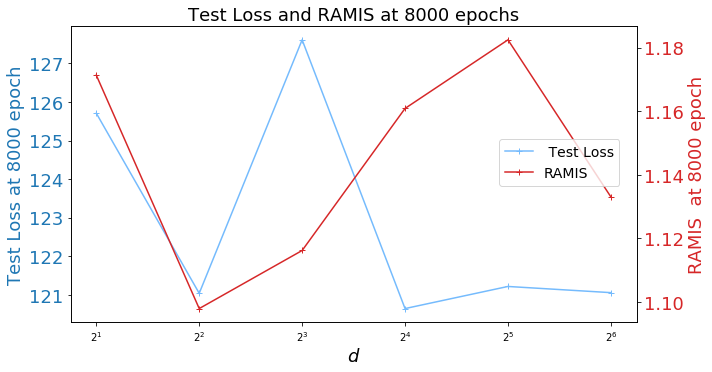}
\caption[VAE Reconstruction]{Test loss and RAMIS Score of VAE for different latent dimensions}
\label{fig:vae_ramis_loss}
\end{figure}
\begin{table}
\begin{tabular}{lllllll}
\hline
\multicolumn{1}{c}{\textbf{d}} & \multicolumn{1}{c}{\textbf{2}} & \multicolumn{1}{c}{\textbf{4}} & \multicolumn{1}{c}{\textbf{8}} & \multicolumn{1}{c}{\textbf{16}} & \multicolumn{1}{c}{\textbf{32}} & \multicolumn{1}{c}{\textbf{64}} \\ \hline
Initial LR($10^{-3}$)        & 1.07                           & 0.96                           & 1.04                           & 0.94                            & 0.88                            & 0.94                            \\
Train Loss                     & 133.6                          & 127.2                          & 136.3                          & 127.2                           & 128.3                           & 127.4                           \\
Test Loss                      & 125.7                          & 121.0                          & 127.6                          & 120.6                           & \textbf{121.2}                  & 121.1                           \\
RAMIS                          & 1.17                           & 1.10                           & 1.12                           & 1.16                            & \textbf{1.18}                   & 1.13                            \\ \hline
\end{tabular}
\caption{Selected initial learning rates, train loss, test loss and RAMIS scores for VAE at different latent dimensions.}
\label{tab:vae_latent_dimension}
\end{table}

\begin{figure}
\centering
\includegraphics[width=0.52\textwidth]{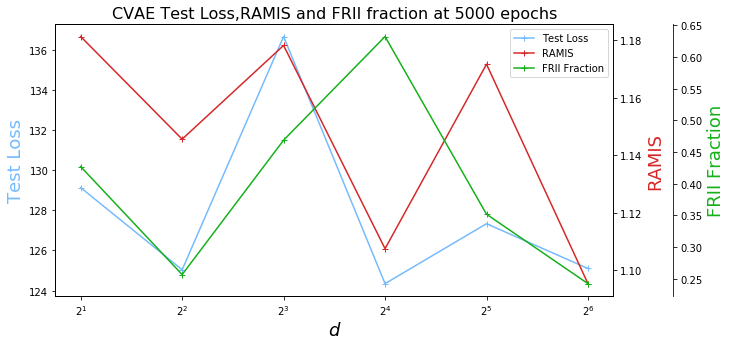}
\caption[VAE Reconstruction]{Test loss, RAMIS score and FRII fraction of the CVAE for different latent dimensions.}
\label{fig:CVAE_latent_dimension}
\end{figure}
\begin{table}
\begin{tabular}{lllllll}
\hline
\multicolumn{1}{c}{\textbf{d}} & \multicolumn{1}{c}{\textbf{2}} & \multicolumn{1}{c}{\textbf{4}} & \multicolumn{1}{c}{\textbf{8}} & \multicolumn{1}{c}{\textbf{16}} & \multicolumn{1}{c}{\textit{\textbf{32}}} & \multicolumn{1}{c}{\textbf{64}} \\ \hline
Initial LR($10^{-3}$)        & 0.90                           & 1.08                           & 0.90                           & 0.94                            & \textit{0.88}                            & 0.96                            \\
Train Loss                     & 136.9                          & 130.9                          & 145.8                          & 133.1                           & \textit{136.0}                           & 132.8                           \\
Test Loss                      & 129.1                          & \textbf{125.0}                 & 136.7                          & \textbf{124.3}                  & \textit{\textbf{127.3}}                  & \textbf{125.1}                  \\
RAMIS                          & \textbf{1.18}                  & 1.14                           & \textbf{1.18}                  & 1.11                            & \textit{\textbf{1.17}}                   & 1.10                            \\
$FRI_{count}$/\%               & 57.5                           & 75.5                           & 53.4                           & 37.1                            & \textit{65.0}                            & 75.9                            \\
$FRII_{count}$/\%              & 37.2                           & 31.8                           & 51.5                           & 61.7                            & \textit{35.6}                            & 24.7                            \\ \hline
\end{tabular}
\caption{Selected initial learning rates, train loss, test loss and RAMIS scores for VAE at different latent dimensions. $FRI_{count}$ and $FRII_{count}$ are the percentage of generated FRIs and FRII correctly classified by the CNN.}
\label{tab:CVAE_latent_dimension}
\end{table}

\subsection{Summary and model selection}

\begin{figure}
\centering
\includegraphics[width=0.5\textwidth]{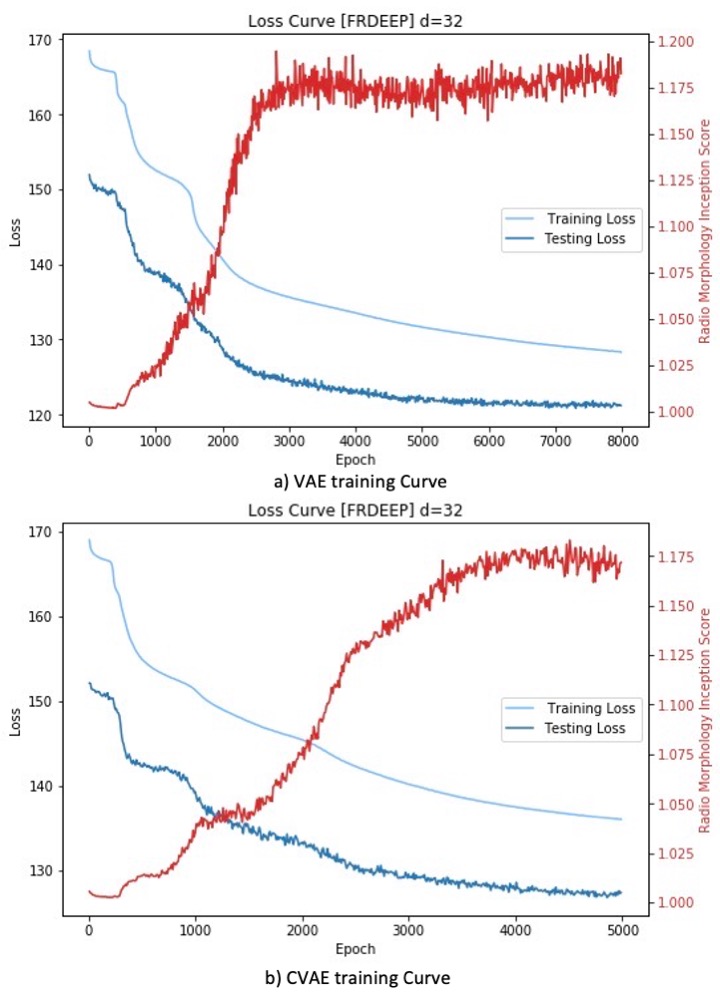}
\caption[VAE Reconstruction]{The VAE and CVAE training curves with the RAMIS curve.}
\label{fig:training_curves}
\end{figure}

We make a model selection using the test loss and the RAMIS score as our main criteria. Table \ref{tab:vae_latent_dimension} shows the metrics for the VAE. As a selection criteria, we make use of the mean RAMIS and mean test loss as a benchmark. Any model with RAMIS larger than the mean RAMIS and with a test loss lower that the mean test loss was selected as being the best performing model. For the VAE we chose $d=32$ at $epoch=8000$ as the best performing model. For the CVAE, only one model conformed to our criteria at $d=32$ with a test loss of $127.33$ (which is lower than the mean $127.93$) with a RAMIS of 1.17 (higher than the mean which is $1.15$). We settled on $d=32$ at $epoch=5000$ for our selected CVAE model. Figure~\ref{fig:training_curves} shows both the training curves for the VAE and CVAE. Samples of images generated by the trained models are presented in Appendix~\ref{app:images}. 

\section{Results}
\label{sec:results}

\subsection{Image Reconstruction}
\label{sec:recon}

\begin{figure*}
\centering
\includegraphics[width=1.0\textwidth]{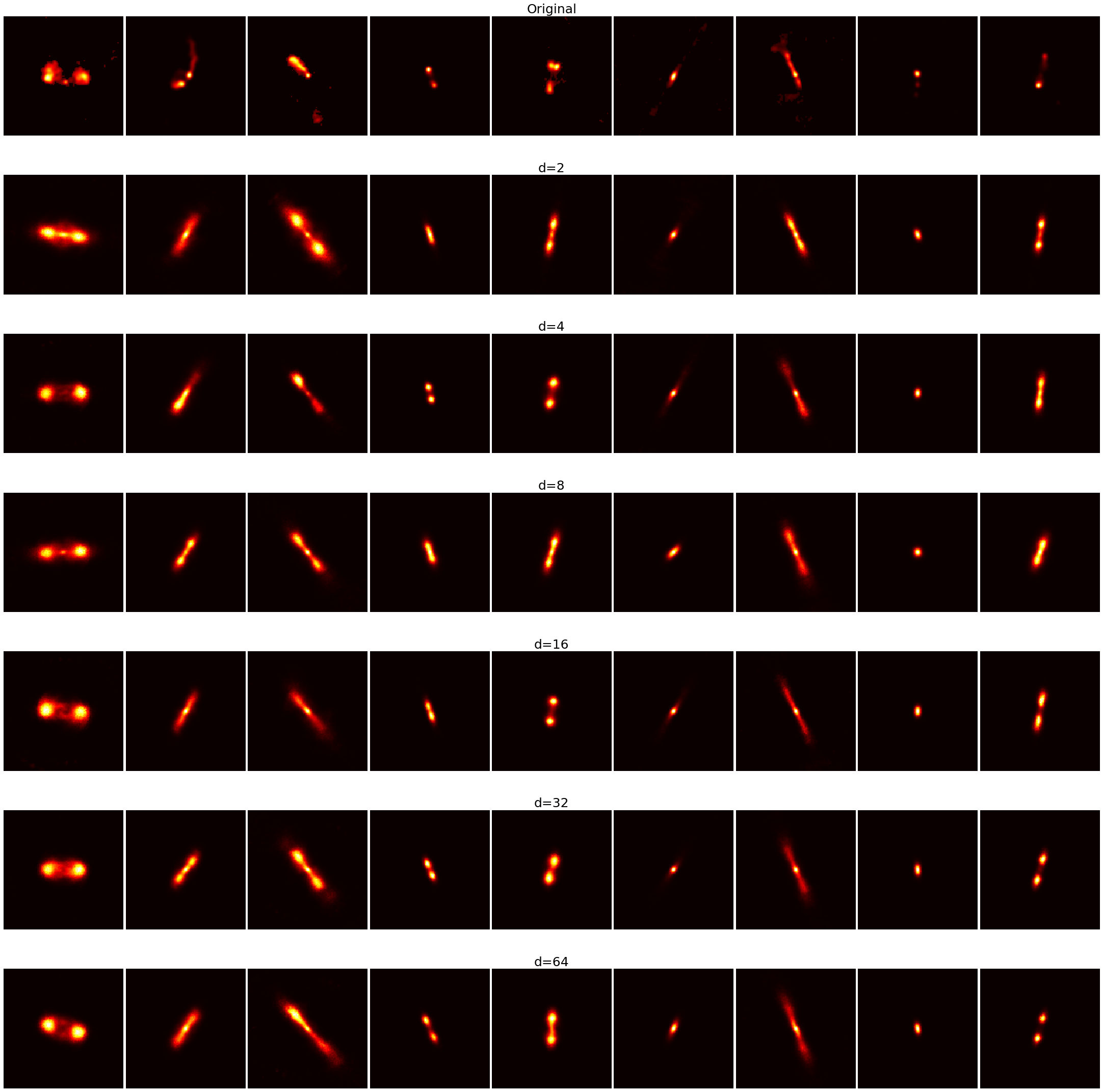}
\caption[VAE Reconstruction for varying $d$ ]{Original and reconstructed images at different latent dimensions. The 1st row shows the original images from the FRDEEP training set. Rows 2 through 6 show the reconstructed images for $d=2,4,8,16,32$ and $64$.}
\label{fig:reconstructedimagemain}
\end{figure*}

To qualitatively understand the general ability of the unsupervised VAE model, a selection of images from the training set were fed into the encoder and reconstructed images generated from those latent coordinates were plotted for each latent dimension, see Figure~\ref{fig:reconstructedimagemain}. As the generation is a stochastic process, it is not expected that the output images will appear identical to the corresponding input, however they should appear similar. In each case the image shown is generated at the minimum test loss and maximum RAMIS score for each model. All 6 latent dimensions were able to reconstruct the radio sources, however some dimensions reproduce images that have structures closer to those of the original images, for example at $d=4$, $d=16$ and $d=32$ the VAE is able to reproduce the asymmetry of the sources: Source~3, which is asymmetric with one lobe brighter than the other, can partially be reconstructed with $d=4$ and $d=16$. The FRI and FRII division can also be reproduced correctly. Source~1, which is an FRII, is reproduced as a radio source with lobes brighter than the core; Source~7, which is an FRI, is reconstructed as a radio source with a bright core and low brightness lobes. However for some latent dimensions, such as $d=8$, we observe that for sources like Source~5, the sources are reconstructed as triple sources while the input image is a double radio source. Another limitation of the system is its inability to reproduce bent structures: Source~2, which is a bent source, is reconstructed as a straight source. This is assumed to be a consequence of having only a small number of bent sources present in the training data. Finally, one of the major drawbacks of the VAE is the `blurriness' of the generated images. This is a known constraint of VAEs \citep{larsen2015} and impacts on the RAMIS score as we have seen in Section~\ref{sec:ramis_performance}. 

\begin{figure*}
\centering
\includegraphics[width=0.9\textwidth]{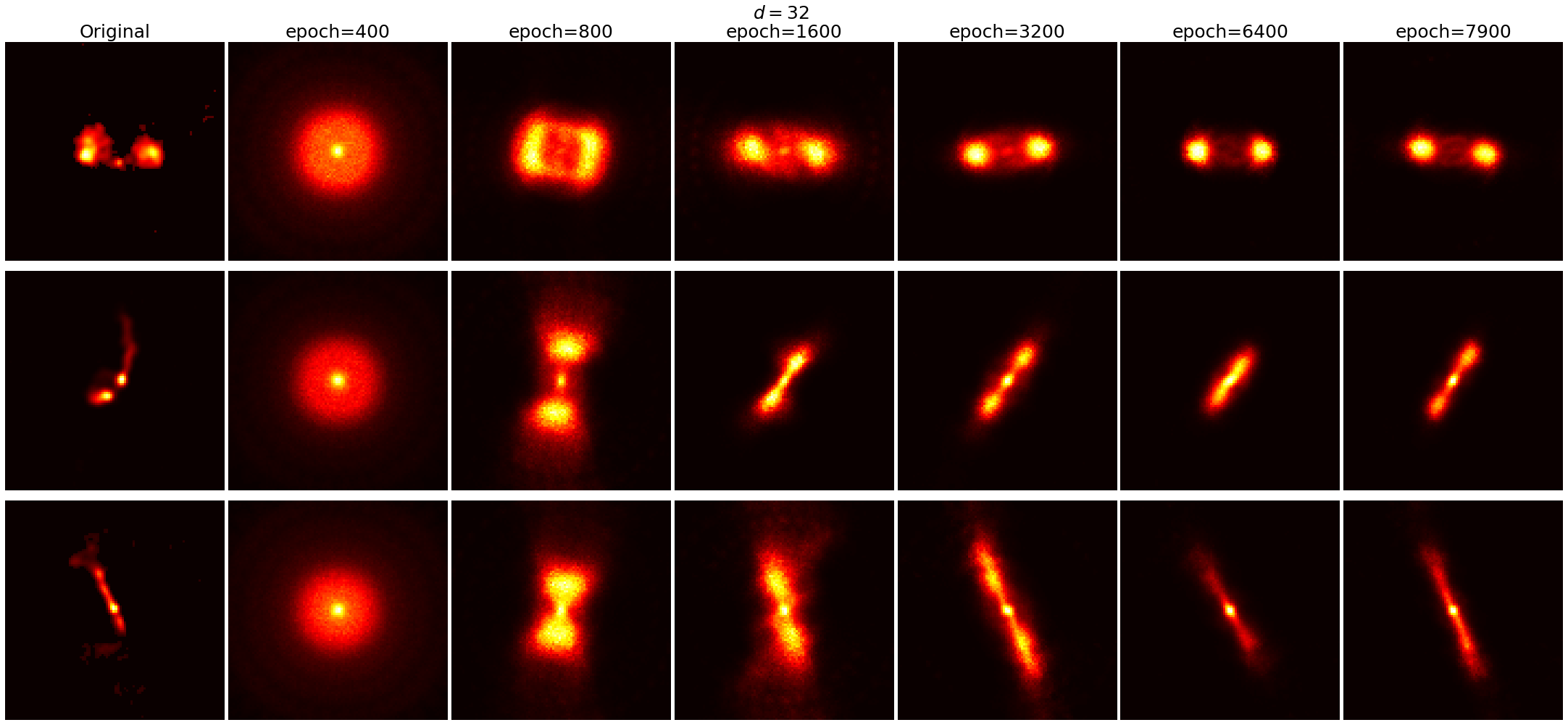}
\caption{Sources generated as a function of training epoch from a model with $d=32$. The first column shows the original image from the training set that is used to define the point in latent space from which the synthetic images are generated.}
\label{fig:reconstructed_d8}
\end{figure*}

Figure~\ref{fig:reconstructed_d8} shows the reconstructed images from the unsupervised VAE with $d=32$ for three different sources from the training set. We can see that the model initially learns the bright central core of the galaxy before learning the extended structure. The orientation of the source is only learned towards the end of the training process.

For the conditional VAE, we can use an input image from the training data set to specify a point in latent space and then choose to generate synthetic images as either FRI or FRII. 

\subsection{Noise Analysis}

As covered in Section~\ref{sec:image_aug}, the training and testing images have been sigma clipped. This resulted in all pixel values below a given threshold being set to zero and created a gap in pixel values between 0 and 0.0039. This is equivalent to approximately $95 \%$ of the pixels being set to zero. Generating images with zero-valued pixels is difficult for machine learning algorithms and they instead assign an infinitely small value to these regions in order to mimic the zeroing process. 
    
\begin{figure*}
    \centering
    \includegraphics[width=0.75\textwidth]{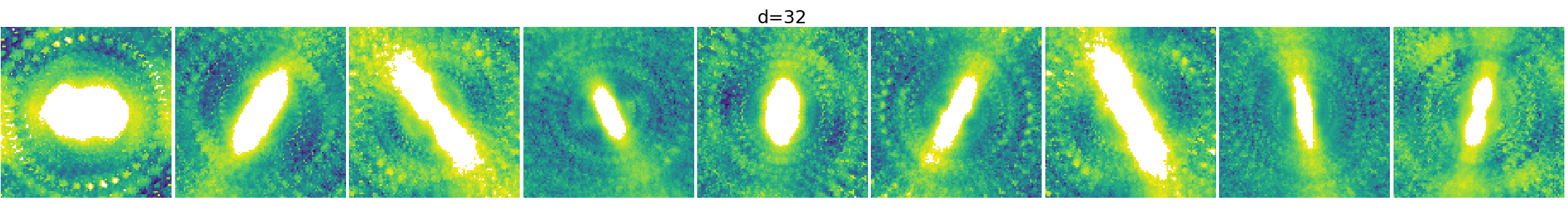}
    \caption{Image of low level noise structures in generated images from a model with $d=32$.}
    \label{fig:noise_imshow}
\end{figure*}

By saturating the generated images at the 95th pixel percentile and using a log-scale to visualise the data we can observe that artifacts are present in the generated images at a low level. These are illustrated in Figure~\ref{fig:noise_imshow}. These artifacts appear as concentric ring-like structures around the generated sources.

The distribution of pixel values in these regions varies between models with different latent dimensions and we suggest that this distribution can be used as an indicator for the performance of the VAE. While zeroing the pixels is computationally difficult, attaining very small values close to zero is an indicator of good network performance. This was evident for models that became stuck in local minima during the training process where the percentage of pixels with values $<5\times10^{-5}$ was significantly higher than those which converged to a global minimum.

\subsection{=2 latent space of the VAE}
\label{sec:2dsom}

\begin{figure*}
    \centering
    \includegraphics[width=0.90\textwidth]{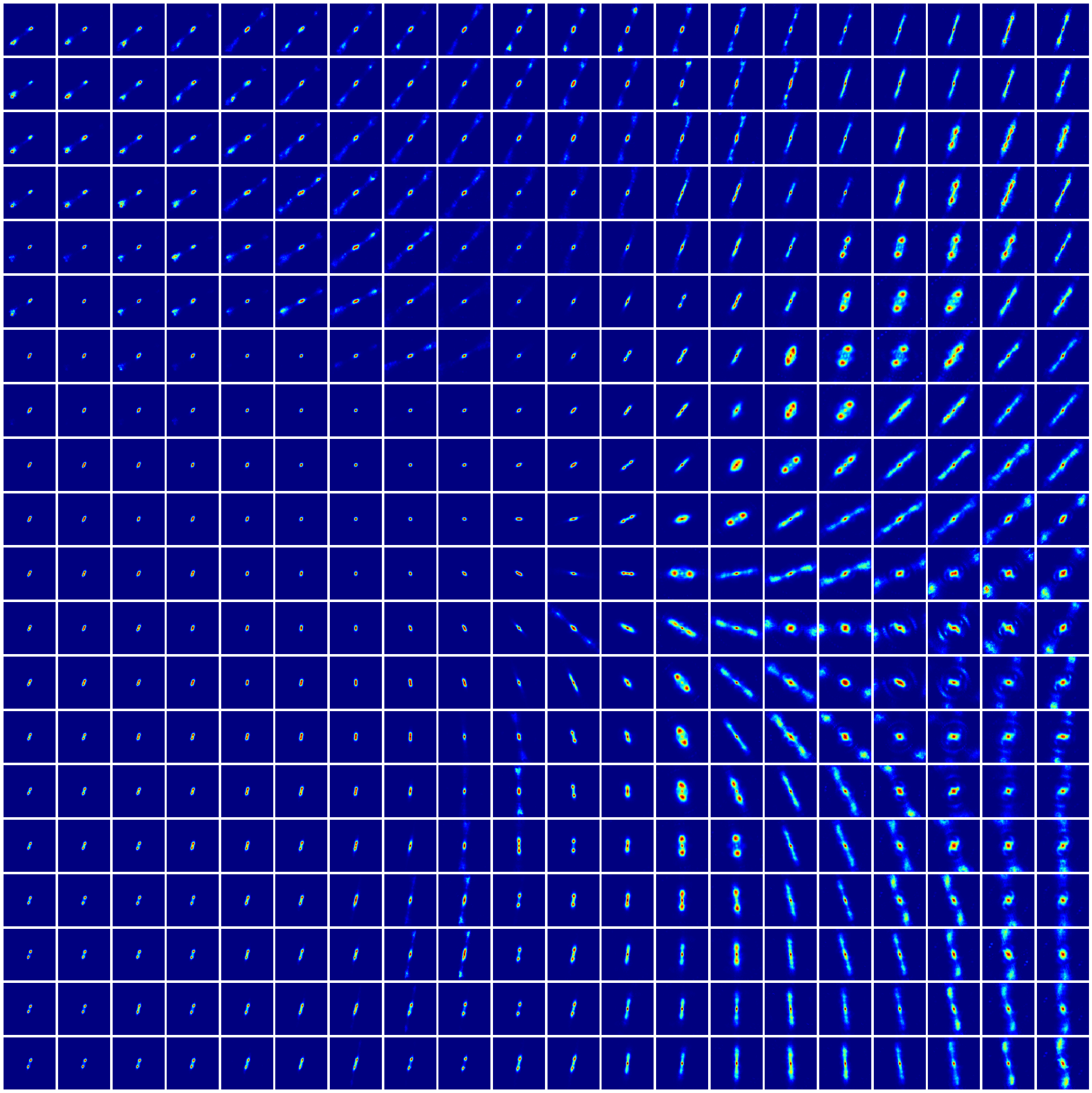}
    \caption[Self Organizing Map of the Variational Autoencoder]{2D Latent mapping of the VAE}
    \label{fig:vae-som}
\end{figure*}

Although the lower dimensionality of the $d=2$ latent space produces sub-optimal generated images, it does enable us to visualise the mapping of the latent variable, $z$, to the output images. Analyzing the latent space can also be useful to visualise the VAEs' ability to separate source characteristics in the 2D latent space based on their morphology. To do this we sample points from the $d=2$ latent space between $-4.0 < z < 4.0$ in steps of 0.4 along the two dimensions of the $d=2$ model and output the images generated at each point. Figure \ref{fig:vae-som} shows the organization of the latent space. 

It can be seen that there is a clear separation between point-like and extended sources in the latent space with point sources being generated at the origin of the latent space. This should be considered when generating sources from the VAE as latent points close to the origin should not be used as these would generate unresolved sources. We also note that moving around the latent space in a clockwise direction results in a change in the orientation of the source, whereby horizontally oriented sources lie towards the latent line $z_1 = 0$. 

For latent points with $z>3$, it can be seen that part of the structure of the generated radio source overlaps the maximum $100\times100$ pixel image extent and, while large extended sources are spread around the latent space, compact double radio sources are concentrated towards the bottom left corner of the space. Towards the top left of the space we find uncentred sources. These sources appear to have one lobe centered at the centre of the image with the other lobe appearing on the lower left quarter of the image.

In the same manner as was illustrated for the $d=2$ latent space of the MNIST data set \citep[][Appendix~2]{kingma2015} we can see that the latent variable, $z$, transitions smoothly between different morphologies represented in the training set, even when those morphologies do not necessarily correspond to an equivalent physical continuum. Whilst this may align with observations of hybrid morphologies in some radio galaxies \citep[e.g.][]{MiraghaeiBest17,Mingo2019RevisitingLoTSS}, it should be considered carefully when generating synthetic examples for data augmentation as the inclusion of too many intermediate morphologies may bias the model performance when applied to real data.

\subsection{Class balance in synthetic source populations}
\label{sec:frcount}

In the case of the conditional VAE, the generated sources were used to evaluate an additional performance metric designed to measure the degree of class separation achieved by the conditional generator. This was calculated by using the classifier described in Section~\ref{sec:ramis_Score} to classify a population of synthetic sources generated by the model with an input specification of a 50:50 FRI:FRII class balance and a uniform random sampling of the latent space in each case. The distributions of classifications are shown in Table~\ref{tab:CVAE_latent_dimension} for each different latent dimension. From Table~\ref{tab:CVAE_latent_dimension}, it can be seen that although a balanced sample was specified at the input to the generator, the resulting synthetic population was found to be imbalanced by the external classifier. More specifically all models were seen to produce an excess of FRI-type sources compared to FRIIs, with the exception of $d=16$. 

If, instead of using a uniform distribution, the latent space is sampled randomly from the prior distribution, $\mathcal{N}(0,1)$, this behaviour is reversed and the synthetic population is found to be dominated by objects classified as FRII-type sources; however, we note that this is likely due to the fact that sources generated from the latent volume around the origin are predominantly compact, comprising unresolved and only marginally resolved objects, and the classifier is biased towards classifying these objects as FRII sources. This effect is also seen when sampling from the prior distribution over the latent space of the VAE. A selection of sources generated by randomly sampling from the prior distribution for the VAE is shown in Figure~\ref{fig:sample vae 1}.

\subsection{CVAE - Class evaluation}

\begin{figure}
\centering
\includegraphics[width=0.5\textwidth]{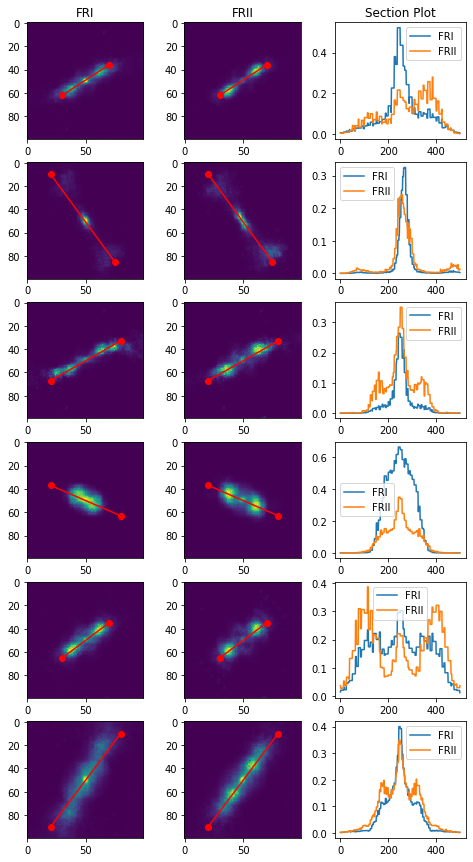}
\caption[VAE Reconstruction]{Comparison of FRI and FRII sources. Column 1 shoes the FRI sources, Column 2 the corresponding FRII source and Column 3 shows the cross sectional plot across the jet axis}
\label{fig:class_evaluation}
\end{figure}

To qualitatively evaluate the CVAE's ability to produce sources with FRI and FRII characteristic, FRI and FRII sources originating from the same latent coordinates were generated The sources were generated from the model with $d=32$ obtained at $epoch=5000$. The cross-sectional profile across the principal axis of the FRI source was compared with that of the FRII. As such two versions (FRI and FRII) of the "same" source could be generated and compared. Figure~\ref{fig:class_evaluation} shows 6 selected sources where the cross-section along the principal axis have been plotted. For each pair of generated sources the main distinctive feature lied in the lobe brightness, generated FRII sources had lobes brighter than the core while for FRIs, the pixel intensity decreases as we move away from the core which in most cases are brighter than the lobe. This is inline with the definition of FRIs and FRIIs, and shows that CVAEs can generate sources with distinctive FRI/II morphological characteristics.

\section{Discussion}
\label{sec:discussion}

Previous work on the generation of radio galaxy images has been undertaken by \cite{ma2019}. In that work the authors used a conditional VAE that used convolutional layers in both the encoder and decoder, which is a significantly different architecture to the fully-connected network presented here. They evaluated their network based on standard classification scores (precision, recall, F1-score) using the classifier defined in \cite{ma2019classifier}, which itself was trained on augmented images produced using a VAE. As already noted in Section~\ref{sec:ramis_performance}, the data pre-processing in \cite{ma2019} involved clipping the training images to $40\times40$\,pixels, which would have impacted significantly on the RAMIS evaluation measure defined in this work and which we suggest may have a disproportionately large effect on the generation of FRII galaxies, which were noted to achieve poorer performance metrics than FRIs in the work of \cite{ma2018}. 

The range of latent space dimensionalities considered by \cite{ma2019} was significantly larger than in this work, with models up to $d=500$ being trained. However their conclusion is inline with the results of this work that find a relatively low latent space dimensionality is preferred. \cite{ma2018} do not explicitly state their preferred dimensionality, but from their Figure~3 it appears to be in the same range as proposed here. 

A further interesting observation in the work of \cite{ma2019} was that the generated images contained two artifacts that are not seen in the generated images from this work. The first of these was described as pseudo-structure, particularly in the generated FRII images, and the second was the presence of a grid structure, or pseudo-texture, overlying the images. \cite{ma2019} attributed this structure to a bias from the mean square error (MSE) loss used to construct those images, equivalent to Equation~\ref{eq:term2}. However, we suggest that it may be a consequence of the convolutional layers used in their network. Chequerboard artifacts in generative algorithms that employ convolutional and specifically deconvolutional layers are a known issue. These artifacts arise from kernel overlap in the deconvolution steps of the decoder \citep{odena2016deconvolution}. Although they can be minimised by the use of deconvolution layers with stride 1 this is typically only used in the final layer of a convolutional decoder and artifacts that have been produced in earlier layers with larger stride steps can still be present. Alternatively, such chequerboard effects can also be mitigated through the use of up-sampling \cite[see e.g.][]{spindler2020}. Other high frequency artifacts are also thought to be caused by the use of max-pooling to subsample the output from convolutional layers in the encoder \citep[e.g.][]{maxpoolerrors}; however, we note that \cite{ma2019} do not use max-pooling to subsample, employing an average pooling approach instead.

While the VAE and CVAE presented in this work can clearly be used to generate realistic radio sources, we note that there remain a number of limitations to this method that should be addressed in future work. The first of these is that VAEs and CVAEs tend to produce blurry images, and while an ideal generative system should generate radio sources with resolution similar to the training set, the resolution of the generated images produced here appears lower than that of the training set, i.e. FIRST. As we have demonstrated, the effect of this blurring will effect performance based on the RAMIS evaluation method introduced in this work. As an alternative to the MSE loss, \cite{ma2019} also used a pixel-wise cross entropy (PCE) loss, which they propose enabled finer structures to be generated in their output images. Another possibility for addressing this resolution issue is to introduce a discriminative network after the VAE identifying input images as real or fake. This approach is known as a VAE-GAN \citep{vaegan}. 

A second limitation is that VAEs cannot reproduce the sigma clipping applied in data pre-processing. In principle this can be remedied by applying a post-processing sigma clipping to the output images, but future applications should also address the nature of the systematic artifacts that appear in this noise.

A final point of note is that the generator is biased towards the creation FRI radio galaxies. With the exception of models with a latent space dimensionality of $d=16$, the CVAE creates a higher number of FRIs compared to FRIIs when the output images are passed through a classifier. Most of these mis-classified FRIIs had an X-shaped morphology or were double-double sources. Compact FRIIs with smaller angular extent were correctly generated and classified while those with large angular extent were generated with unclear morphologies. This is similar to the performance mis-match seen by \cite{ma2018} for their FRI/II populations.

\section{Conclusions}
\label{sec:conclusion}

In this work we have demonstrated the use of generative machine learning methods to simulate realistic radio galaxies. We present results from both an unsupervised variational autoencoder and a conditional variational autoencoder. The networks were trained using sources from the FIRST radio survey and produce radio sources with FRI and FRII morphologies. Furthermore we have presented a quantitative method for evaluating the performance of these generative models in radio astronomy, formulated as the radio morphology inception score (RAMIS).

Using both the RAMIS as a quantitative measure and by inspecting the radio sources, we found that VAEs could be used as a method for the generation of relatistic radio sources. We found that the lowest model loss was obtained at a latent dimension of $d=32$ with a RAMIS of $1.175$. However, we also found that the VAE could correctly construct asymmetry in the radio sources at the lower latent dimension, $d=4$. 
    
We also investigated the mapping of the latent space to output images, this was done by visualising the generated sources plotted from different latent points in the $d=2$ latent space. We identified a ystematic distribution of morphologies in the latent space with extended radio sources separating themselves from point like sources. We also investigated the class balance in the generative source population for the CVAE, illustrating the difference in outcomes when sampling from the latent space in different ways. We suggest that the implications of this investigation can be used if some control is needed over the generation of synthetic radio sources but caution that it also highlights the potential for bias when being used to augment data sets for training other models.

\section*{Acknowledgements}

The authors thank the reviewer for their comments, which significantly improved this paper. The authors would like to thank Alex Shestapoloff and Brooks Paige from the Alan Turing Institute and Tingting Mu from the University of Manchester for early discussions that informed the course of this work. DJB gratefully acknowledges support from STFC and the Newton Fund through the DARA Big Data program under grant ST/R001898/1. AMS gratefully acknowledges support from an Alan Turing Institute AI Fellowship EP/V030302/1. MB gratefully acknowledges support from the University of Manchester STFC CDT in Data Intensive Science, grant number ST/P006795/1. FP gratefully acknowledges support from STFC and IBM through the iCASE studentship ST/P006795/1. 

\section*{Data and source code Availability}

Trained models from this work are publicly available on Zenodo (DOI:10.5281/zenodo.4456165). All codes are available on github: \url{https://github.com/joshen1307/RAGA}. The FRDEEP dataset used in this work is publicly available from Zenodo (DOI:10.5281/zenodo.4255826) and should be cited as \cite{tang2019}.



\bibliographystyle{mnras}
\bibliography{vae_paper} 


\appendix
\section{Images}
\label{app:images}
\begin{figure}
\centering
\includegraphics[width=0.45\textwidth]{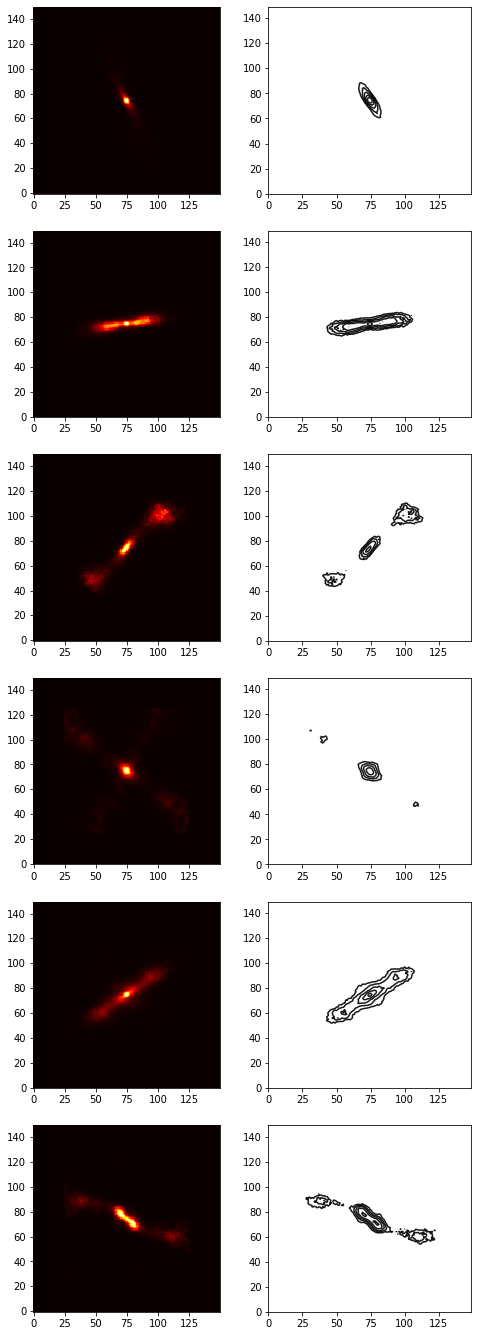}
\caption[FR1 Set 1]{Generated FRI Sample Set 1}
\label{fig:fr1_set1}
\end{figure}

We here present samples of generated sources using the trained VAE and CVAE. 
\begin{enumerate}
    \item Figure \ref{fig:fr1_set1}  and \ref{fig:fr1_set4} : Generated FRI sources from the CVAE that were classified as FRI by the CNN classifier.
    \item Figure \ref{fig:fr2_set1}  and \ref{fig:fr2_set4} : Generated FRII sources from the CVAE that were classified as FRII by the CNN classifier.
    \item Figure \ref{fig:sample vae 1} : 100 generated sources from the unsupervised VAE for $d=32$ where $z$ is sampled randomly from the prior $N(0,1)$.
\end{enumerate}

\begin{figure}
\centering
\includegraphics[width=0.45\textwidth]{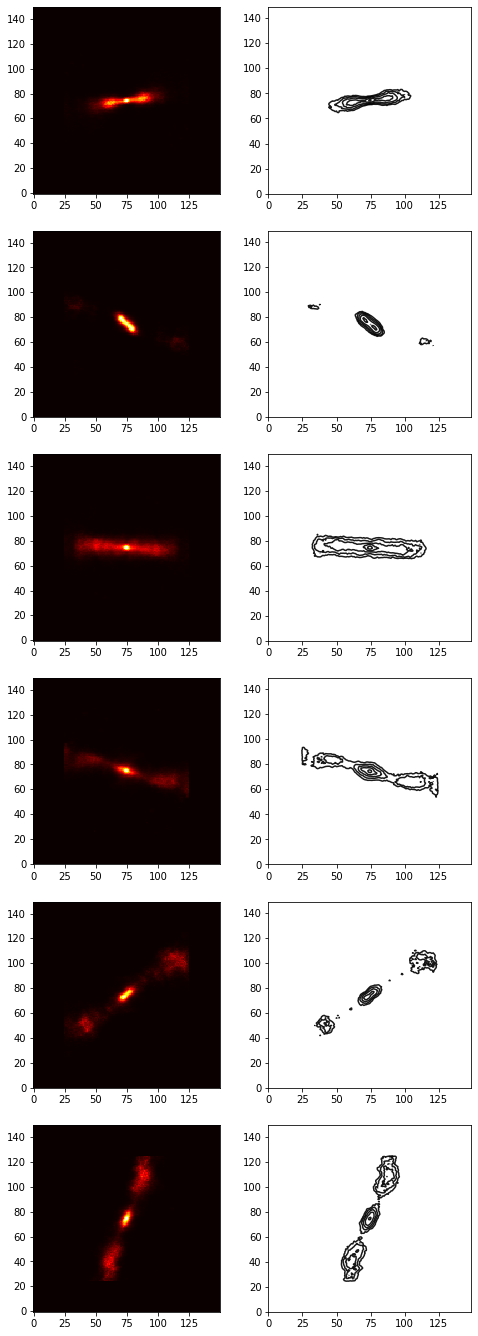}
\caption[FR1 Set 2]{Generated FRI Sample Set 2}
\label{fig:fr1_set2}
\end{figure}

\begin{figure}
\centering
\includegraphics[width=0.45\textwidth]{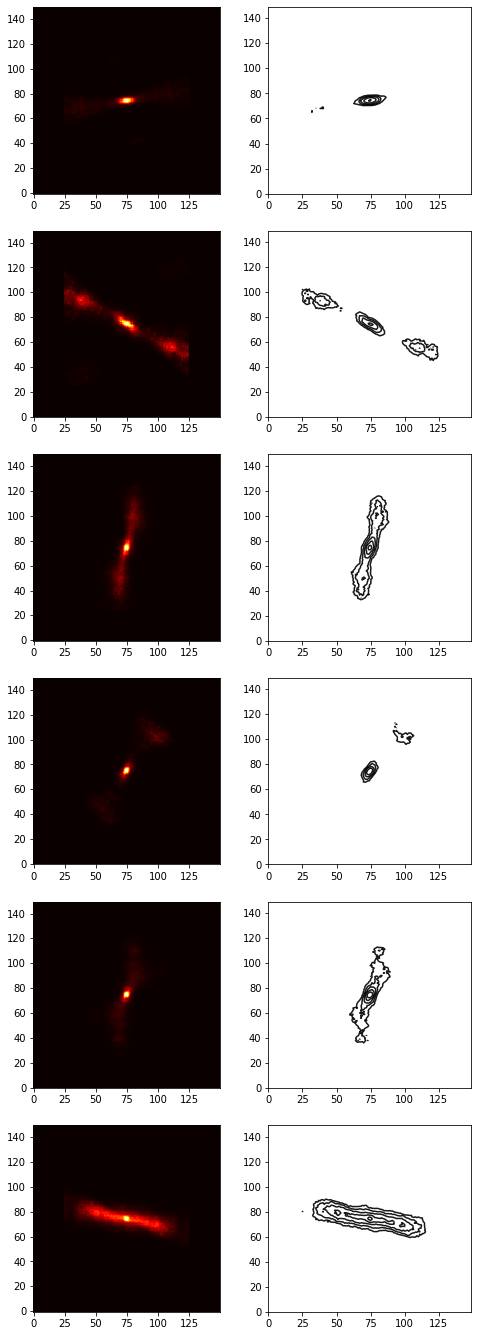}
\caption[FR1 Set 3]{Generated FRI Sample Set 3}
\label{fig:fr1_set3}
\end{figure}

\begin{figure}
\centering
\includegraphics[width=0.45\textwidth]{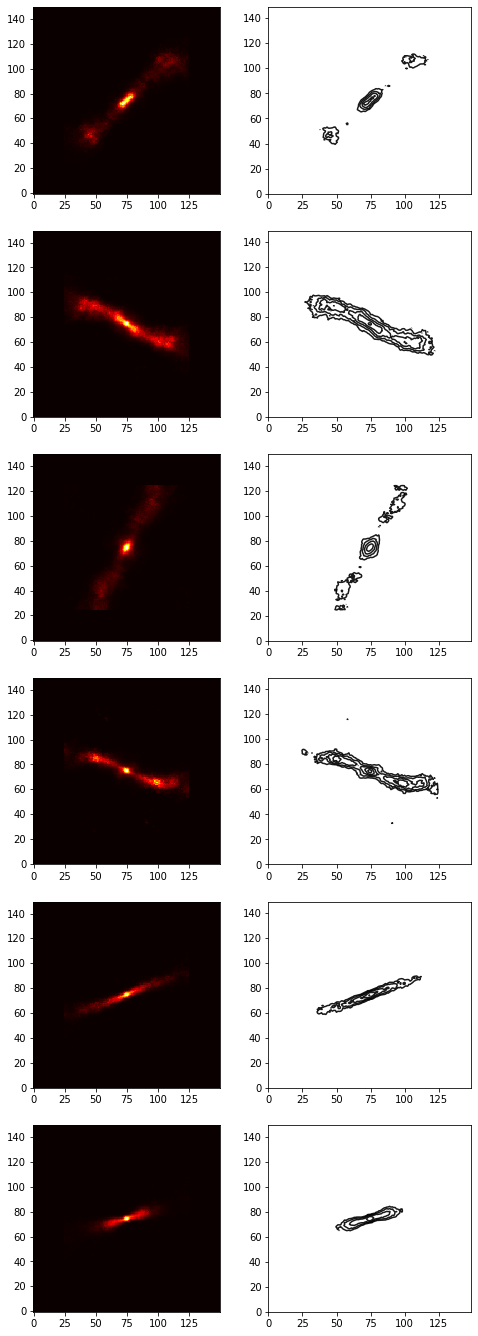}
\caption[FR1 Set 4]{Generated FRI Sample Set 4}
\label{fig:fr1_set4}
\end{figure}

\begin{figure}
\centering
\includegraphics[width=0.45\textwidth]{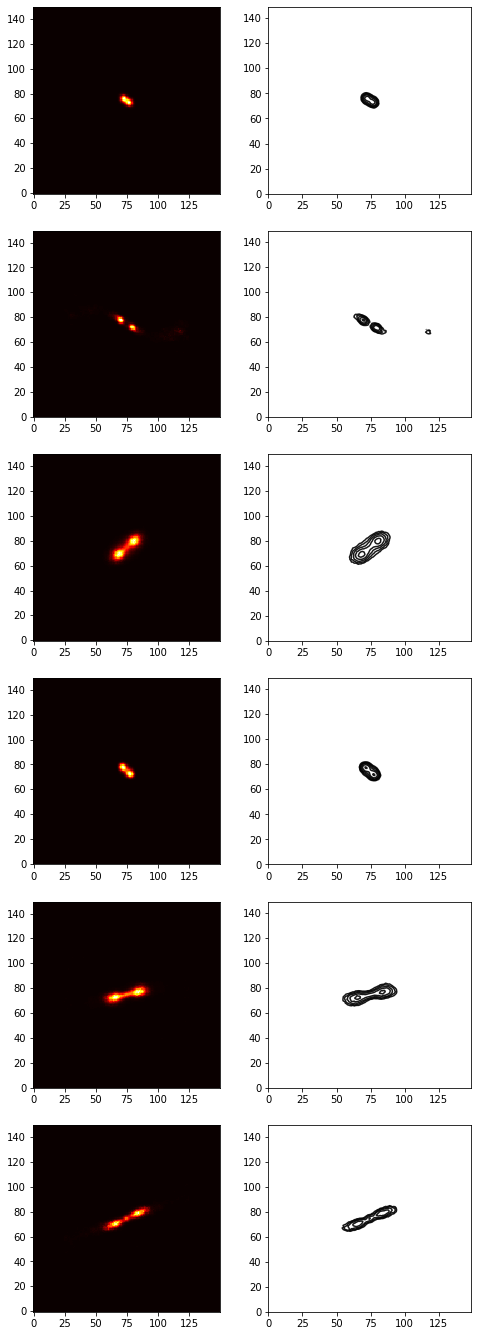}
\caption[FR2 Set 1]{Generated FRII Sample Set 1}
\label{fig:fr2_set1}
\end{figure}

\begin{figure}
\centering
\includegraphics[width=0.45\textwidth]{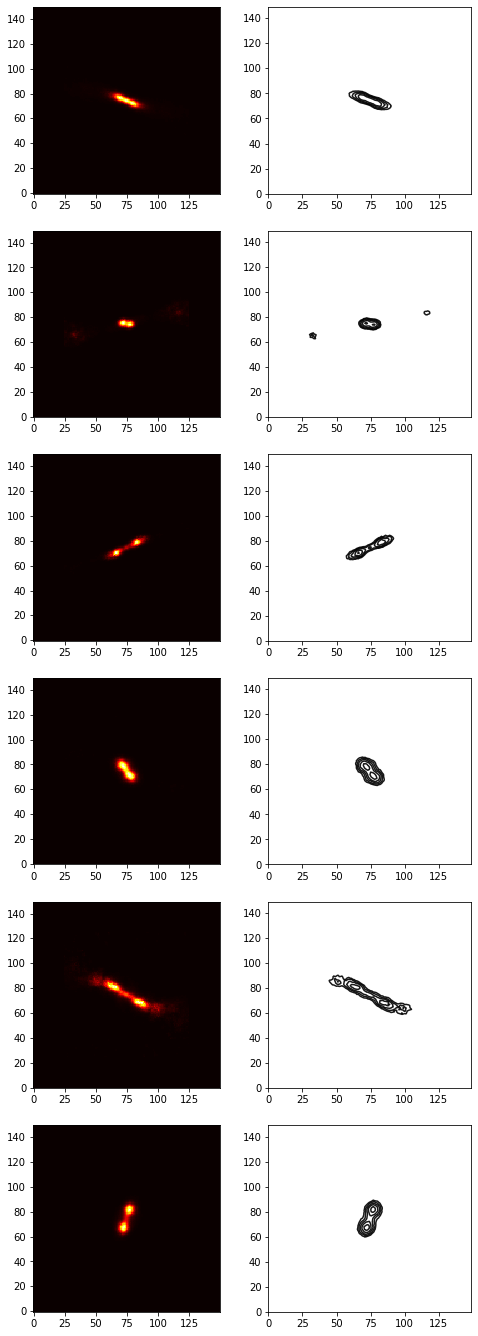}
\caption[FR2 Set 2]{Generated FRII Sample Set 2}
\label{fig:fr2_set2}
\end{figure}

\begin{figure}
\centering
\includegraphics[width=0.45\textwidth]{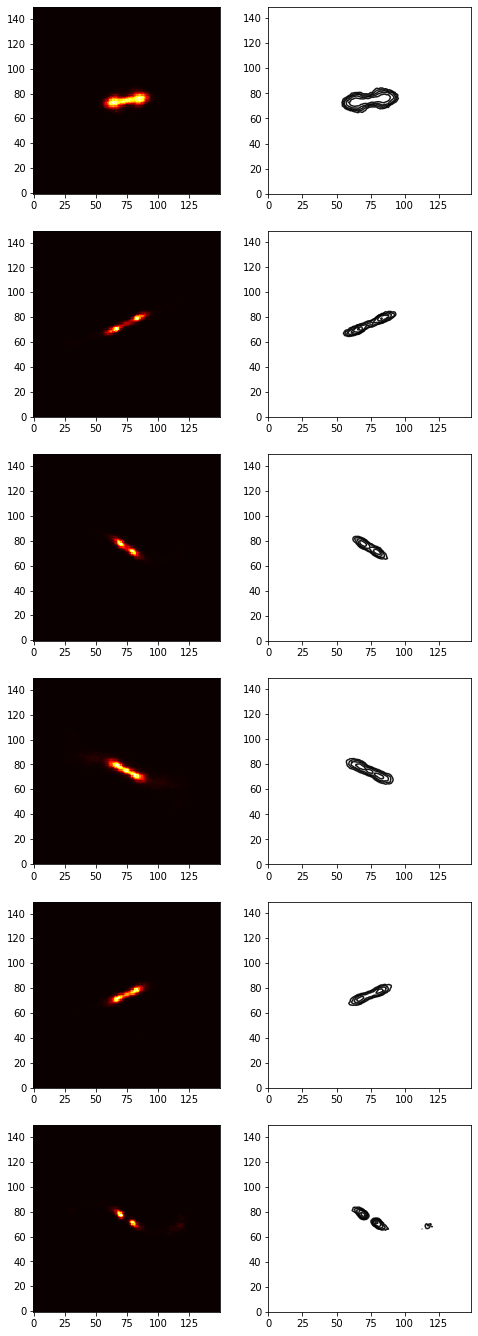}
\caption[FR2 Set 3]{Generated FRII Sample Set 3}
\label{fig:fr2_set3}
\end{figure}

\begin{figure}
\centering
\includegraphics[width=0.45\textwidth]{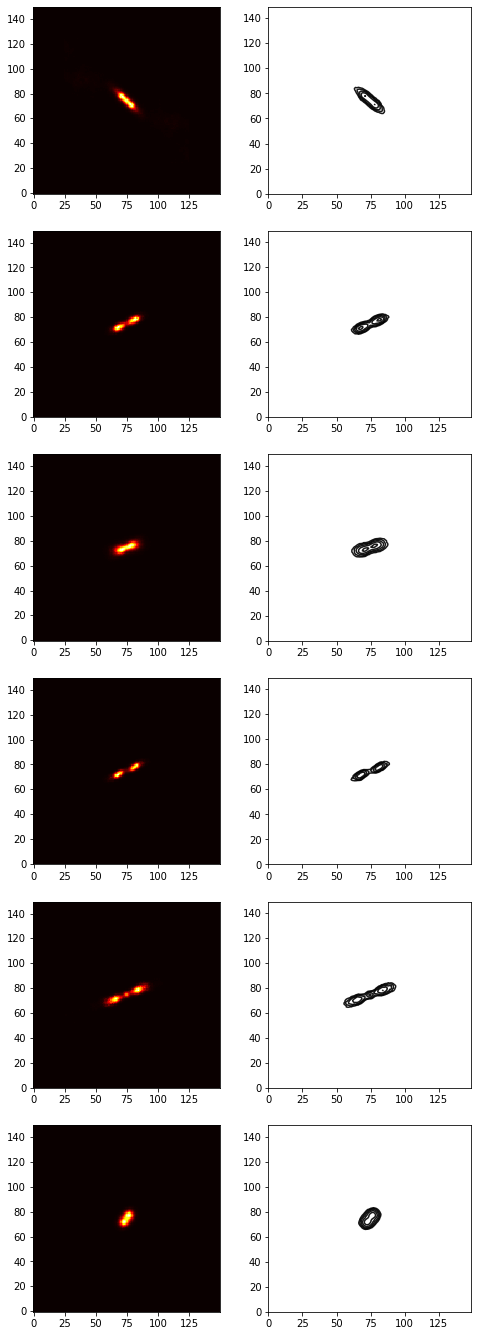}
\caption[FR2 Set 4]{Generated FRII Sample Set 4}
\label{fig:fr2_set4}
\end{figure}

\begin{figure*}
    \centering
    \includegraphics[width=0.9\textwidth]{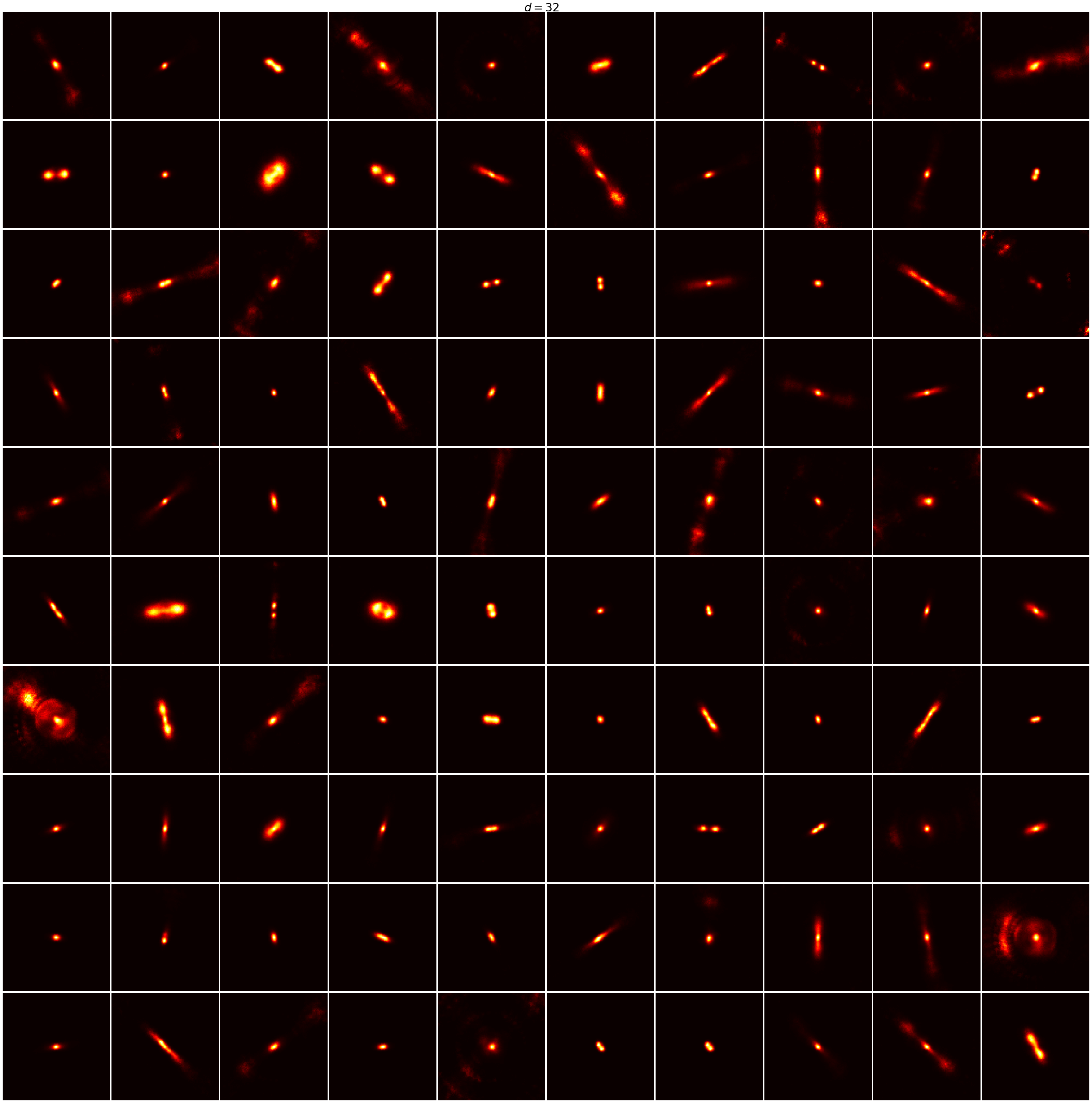}
    \caption{Generated sources from the unsupervised VAE for $d=32$ where $z$ is sampled randomly from the prior $N(0,1)$.}
    \label{fig:sample vae 1}
\end{figure*}


\bsp	
\label{lastpage}
\end{document}